\documentclass[times, preprint]{aastex63}
\definecolor{citeE}{rgb}{0.1, 0.5, 0.9}
\definecolor{citegrey}{rgb}{0.45, 0.5, 0.55}
\definecolor{darkblue}{rgb}{0., 0., 0.8}
\hypersetup{linkcolor=red,citecolor=citeE, filecolor=cyan, urlcolor=darkblue}
\submitjournal{the Astrophysical Journal}
\shorttitle{Solar minimum and Martian gravity waves}
\graphicspath{{./}{./Figures/}} 

\newcommand{\one}[1]{\protected@edef\@tempa{#1}\ul\@tempa}

\begin{document}
\title{Variations of the Martian Thermospheric Gravity Wave Activity \\ during the Recent Solar Minimum as Observed by MAVEN}
\author[0000-0002-2819-2521]{Erdal Y\.I\u g\.It}\affiliation{George Mason University\\4400 University Drive \\Fairfax, VA 22030, USA}
\author{Alexander S. Medvedev}\affiliation{Max Planck Institute for Solar System Research, G\"ottingen, Germany}
\author{Paul Hartogh}\affiliation{Max Planck Institute for Solar System Research, G\"ottingen, Germany}
\begin{abstract}
  Atmospheric gravity  (buoyancy) waves (GWs) are of great importance for the energy and momentum budget of all planetary atmospheres. Propagating upward waves carry energy and momentum from the lower atmosphere to thermospheric altitudes and re-distribute them there. On Mars, GWs dominate the variability of the thermosphere and ionosphere. We provide a comprehensive climatology of Martian thermospheric GW activity at solar minimum (end of Solar Cycle 24) inferred from measurements by Neutral Gas and Ions Mass Spectrometer on board Mars Atmosphere and Volatile EvolutioN (NGIMS/MAVEN). The results are compared and interpreted using a one-dimensional spectral nonlinear GW model. Monthly mean GW activity varies strongly as a function of altitude (150--230 km) between 6-25\%, reaching a maximum at $\sim$170 km. GW activity systematically exhibits a local time variability with nighttime values exceeding those during daytime, in accordance with previous studies. The analysis suggests that the day-night difference is primarily caused by a competition between dissipation due to molecular diffusion and wave growth due to decreasing background density. Thus, convective instability mechanism is likely to play a less important role in limiting GW amplitudes in the upper thermosphere, which explains their local time behavior.     
\end{abstract}

\section{Introduction}

Planetary upper atmospheres and ionospheres are influenced from below by internal waves and from above by solar and geomagnetic processes at various time scales \citep{Kutiev_etal13, Bougher_etal15, Yigit_etal16b, Zurek_etal17, Thiemann_etal18, MedvedevYigit19}. The 11-year solar cycle is known to significantly impact the upper atmosphere of Earth \citep{Chanin07, Vickers_etal14} and Mars \citep{Bougher_etal99, Bougher_etal15, Gonzalez-Galindo_etal15}. The solar cycle is thought to be produced by the conversion of poloidal magnetic fields to azimuthal fields by a self-generated dynamo on the Sun \citep{Dikpati_etal04, Svalgaard_etal05}. A direct manifestation of this dynamical process is the occurrence of active regions and sunspots in the photosphere, which have been counted for a long time to derive a proxy for solar activity. 

The variation of the solar activity influences the thermosphere, especially the plasma properties (densities and temperatures of electrons and ions, critical frequencies), ion-neutral coupling, and composition. These changes ultimately produce large-scale dynamical and thermodynamical response in the upper atmosphere \citep{Yigit18}. During a solar cycle, the ultraviolet (UV, $\lambda \sim 10-380$ nm) portion of the solar spectrum undergoes a significant degree of variability. The UV and extreme UV (EUV) radiation impacts planetary thermospheres primarily via photoabsorption and photoionization processes, which change the thermospheric temperature and density. The Solar Cycle 24 has been characterized by an extended period of low solar activity with an unprecedented minimum of sunspots \citep{Li_etal19, Pesnell20}. From the perspective of atmospheric vertical coupling, it is  convenient to study the upper atmosphere during a solar minimum, because the effect of solar variability on the upper atmosphere is minimal, and the signatures of upward propagating waves are more direct. 

While the upper atmosphere undergoes long-term changes due to varying solar irradiation, the thermosphere is continuously disturbed by atmospheric gravity (buoyancy) waves (GWs) propagating upward from the lower atmosphere. GWs are ubiquitous features of all planetary atmospheres \citep{YigitMedvedev19}. They have been detected in the thermospheres of Venus \citep{Garcia_etal09}, Earth \citep{Forbes_etal16}, Mars \citep{Yigit_etal15b}, Jupiter \citep{Young_etal97, WatkinsCho13}, and Saturn \citep{Muller-Wodarg_etal19}. There is a growing interest in studying them in exoplanetary atmospheres \citep{WatkinsCho10}.  GWs have been unambiguously observed also in the lower atmosphere of the Sun and are thought to contribute to the overall energy budget \citep{Straus_etal08}. Their effects have been most extensively studied in the upper atmosphere of Earth using numerical models \citep{ HickeyCole88, Hickey_etal11, WalterscheidHickey11, LundFritts12,  GavrilovKshevetskii15, GavrilovKshevetskii13, Fritts_etal15a, MiyoshiYigit19, YigitMedvedev17, Lilienthal_etal20, Gavrilov_etal20, Yigit_etal21a}.  Similarly, an increasing number of numerical modeling studies have demonstrated the dynamical and thermodynamical importance of  GWs  in the atmosphere of Mars. One-dimensional linear full-wave modeling study conducted by \citet{Parish_etal09} has suggested that GWs can directly propagate from the troposphere (i.e., lower atmosphere) to the thermosphere and induce density fluctuations comparable to those from the Mars Odyssey aerobraking measurements. Using a nonlinear numerical model, \citet{Jiang_etal21}  have shown that the Martian ionospheric irregularities can be seeded by variations  of the neutral wind, which are largely associated with GWs. Three-dimensional time-dependent Martian general circulation modeling with an implemented subgrid-scale nonlinear GW parameterization has predicted that  the waves significantly drive the atmospheric circulation in the mesosphere and thermosphere \citep{Medvedev_etal11a}, cool the Martian upper atmosphere \citep{MedvedevYigit12}, and facilitate the formation of carbon dioxide ice clouds \citep{Yigit_etal18, Yigit_etal15a}. A one-dimensional modeling study demonstrated that wave-induced temperature fluctuations can enhance the Jeans escape flux on Mars \citep{Walterscheid_etal13}. Overall, theoretical modeling efforts systematically demonstrated the dynamical and thermodynamical importance of GWs for the Martian whole atmosphere system.

 Martian GWs have been systematically observed in the thermosphere by a number of spacecraft at various altitudes in terms of density and temperature fluctuations \citep{Fritts_etal06, Tolson_etal07, Yigit_etal15b, England_etal17, Terada_etal17, Jesch_etal19, Siddle_etal19, Vals_etal19, Li_etal21}, suggesting that they are present there at all altitudes and times. In light of previous numerical modeling predictions, it is likely that much of the observed thermospheric small-scale variability is driven by GWs.

The upper atmospheric temperature, winds, and density change over a solar cycle as a response to the varying energy deposition by the solar UV. Therefore, the associated thermospheric GW activity is expected to vary as well. Terrestrial studies have provided some insight into the wave response to varying solar activity. General circulation modeling studies with parameterized subgrid-scale  GWs have demonstrated that wave amplitudes and associated wave-induced acceleration/deceleration (i.e., wave drag) are stronger when the solar activity is low  \citep{YigitMedvedev10}. Global observations of relative density fluctuations by the CHAMP satellite have provided further evidence for a clear response of the thermospheric GW activity to solar cycle variations \citep{Park_etal14} and qualitatively corroborated the modeling predictions.  \textit{On Mars, the characterization of thermospheric GW activity during the solar cycle is yet to be established.} Our paper contributes to this goal by considering the behavior of GWs at the recent solar minimum.

\begin{figure*}[t!]
  \hspace*{-0.5cm}
  \centering \includegraphics[width=0.45\textwidth]{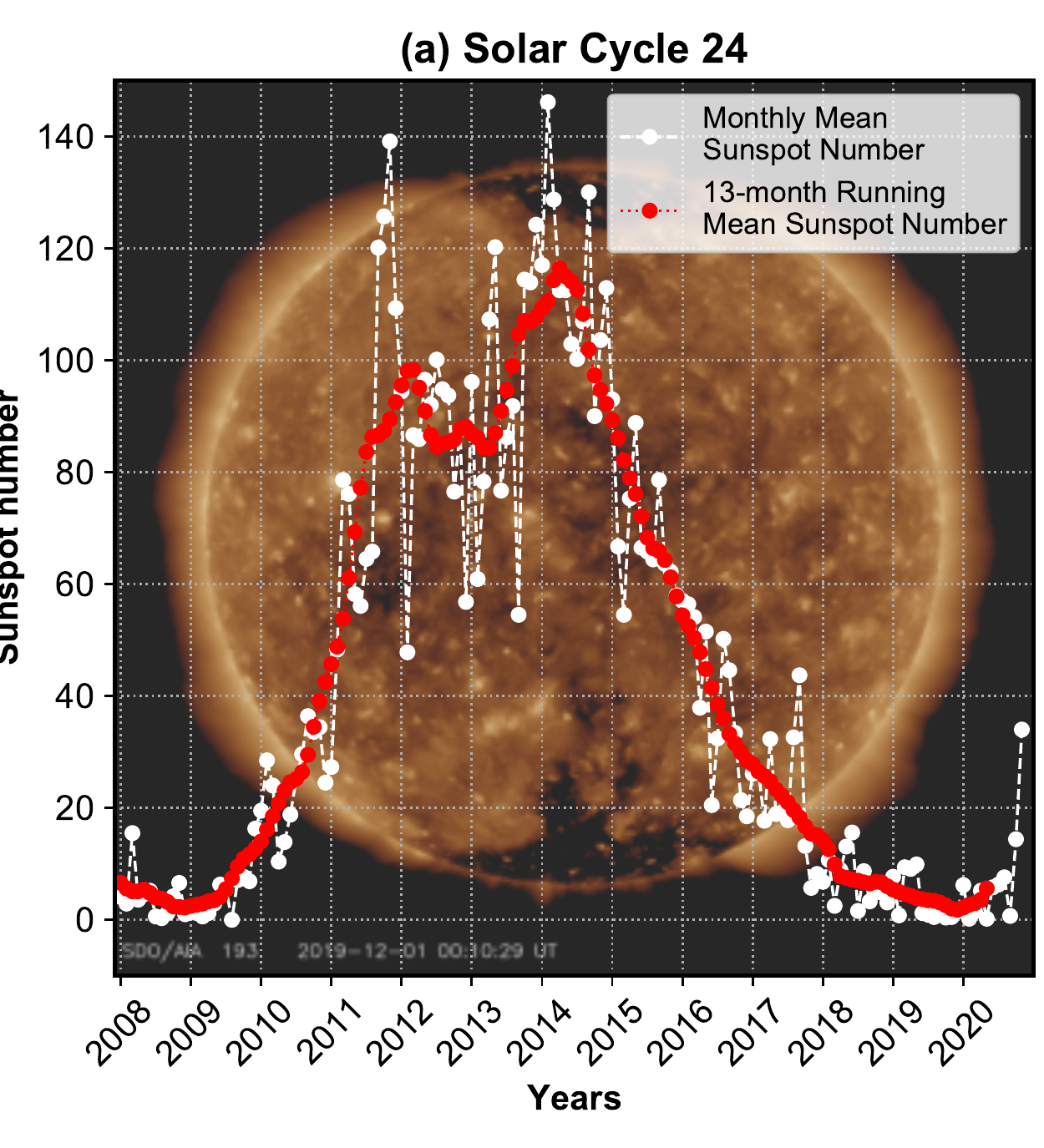}
  \centering \includegraphics[width=0.45\textwidth]{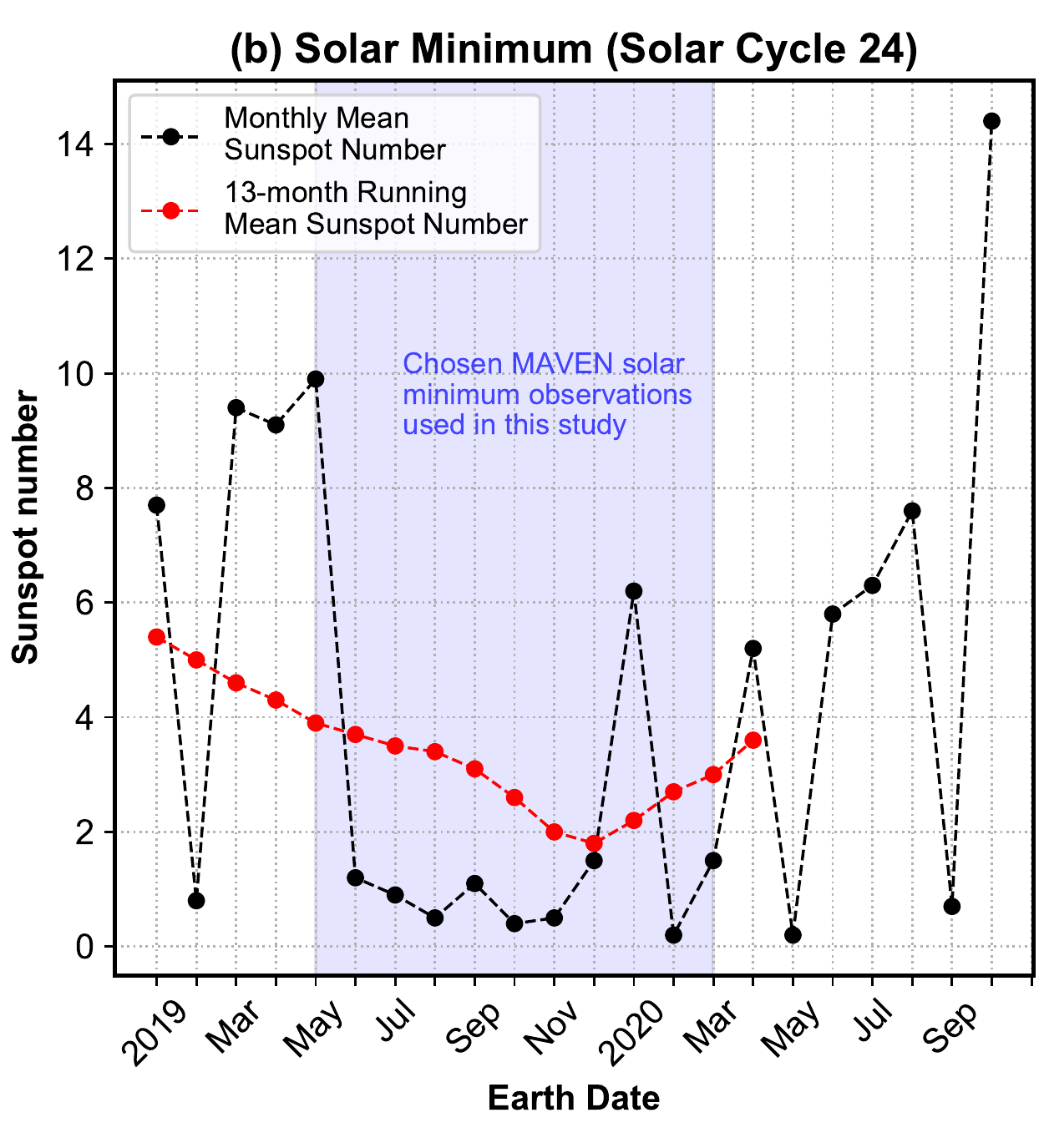}
    \caption{(a) Solar cycle 24 in terms of monthly mean (white) and 13-month running mean (red) international sunspot numbers ($S_n$). The period of the MAVEN data used in this study is shaded in light blue in panel (b) where the solar minimum period is shown in more detail. Data source: WDC-SILSO, Royal Observatory of Belgium, Brussels. The background image in panel a is the image of the Sun taken at $\lambda=193$ \AA~ on 1 December 2019 at 00:10:29 UT by the Atmospheric Imaging Assembly (AIA) instrument on board the Solar Dynamic Observatory.}
    \label{fig:sunspots}
  \end{figure*}

NASA's Mars Atmosphere and Volatile EvolutioN (MAVEN) spacecraft has been providing scientific observations since late-2014 until today \citep{Bougher_etal15b}, therefore fully covering the  recent solar minimum. MAVEN is the first mission fully dedicated to the characterization of the Martian upper atmosphere. It has been extensively used to study various aspects of the Martian ionosphere and thermosphere, such as electron and neutral temperature retrievals \citep{VigrenCui19}, ion-neutral coupling \citep{Mayyasi_etal19}, thermospheric circulation \citep{Benna_etal19},  thermospheric dayside temperature and scale height trends \citep{Bougher_etal17}, and characterization of the 2018 planet-encircling dust storm effects in the thermosphere \citep{Jain_etal20}. MAVEN provides a unique opportunity to characterize GW activity as well. Its usefulness for exploring GW processes has been demonstrated by a number of recent studies \citep{Yigit_etal15b, England_etal17,  Leelavathi_etal20, Li_etal21, Yigit_etal21b}. 
In this paper we determine a detailed climatology of thermospheric GWs around the recent solar minimum that took place around December 2019 and provide a physical interpretation for the observed local time variability using a GW model.

The next section describes the methods and data, including Solar Cycle 24, MAVEN's data coverage, and extraction of gravity waves from MAVEN/NGIMS observations. Results are presented in Section~\ref{sec:results}. Discussion of the observations and the GW model analysis are given in Section~\ref{sec:discussion}. Summary and conclusions are presented in Section \ref{sec:summary-conclusions}.

\section{Methods and Data}
\subsection{Solar Cycle 24 and Solar Minimum}
Solar activity is often characterized  by the number of sunspots, which are cooler ($T\approx 3700$ K) and darker Earth-sized regions of enhanced magnetic activity in the solar photosphere ($T\approx 5800$ K), the visible surface of the Sun \citep{Kilcik_etal11, Kiess_etal14, Yigit_etal18a}. Figure \ref{fig:sunspots}a presents the Solar Cycle 24 as a function of the international sunspot number, $S_n$, shown as monthly mean (black) and 13-month running mean (red), based on the SILSO World Data center at the Royal Observatory of Belgium. The most recent Solar Cycle 24 started in December 2008. The activity was minimal until early 2010, reached a peak in April 2014 with $S_n \sim 116$ (see the smoothed 13-month weighted curve) and ended in December 2019 with $S_n \sim 1.8$. The background image shows the Sun at captured on 1 December 2019 at 00:10:29 UT by the Atmospheric Imaging Assembly (AIA) instrument on board the Solar Dynamic Observatory at $\lambda=193$ \AA, showing the Sun at a representative low minimum activity. MAVEN scientific observations started during the descending phase of the solar cycle, when the solar activity was moderate. In this study, we exploit MAVEN's continuous coverage during the last solar minimum. Specifically, a period of 10 months with very low mean sunspot number is chosen for analysis, as highlighted in Figure \ref{fig:sunspots}b.

\subsection{MAVEN/NGIMS Data}\label{sec:data-analysis}
\begin{deluxetable*}{cccc}\tablenum{1}
\tablecaption{Details of the 1307 MAVEN orbits used in this study.}
\tablewidth{0pt}
\tablehead{
\colhead{Number of } & \colhead{Earth} & \colhead{Martian Date} &\\ 
\colhead{available orbits}     & \colhead{Date}    & \colhead{$L_s$ (Solar longitude)}  & 
}
\startdata
146  & 1-31 May 2019  & 18.6$^\circ$-32.9$^\circ$ &  \\
127  & 1-30 Jun 2019  & 32.9$^\circ$-46.3$^\circ$ &  \\
156  & 1-31 Jul 2019  & 46.4$^\circ$-60.0$^\circ$ &  \\
113  & 1-31 Aug 2019  & 60.1$^\circ$-73.6$^\circ$ &  \\
76   & 1-30 Sep 2019  & 73.7$^\circ$-86.7$^\circ$ &  \\
137 & 1-30 Oct 2019 &  86.8$^\circ$-100.4$^\circ$&  \\
137 & 1-30 Nov 2019 &  100.5$^\circ$-114.1$^\circ$& \\
145 & 1-31 Dec 2019 &  114.2$^\circ$-128.6$^\circ$& \\
146 & 1-31 Jan 2020 &  128.7$^\circ$-143.8$^\circ$& \\
124 & 1-29 Feb 2020 &  143.9$^\circ$-158.8$^\circ$& \\
\enddata
\tablecomments{Orbits with scientifically unusable data have been excluded.}
\label{tb:orbit_data}
\end{deluxetable*}

\begin{figure*}[t!]
  \vspace{-0.1cm}
  \hspace*{-1.cm} \centering
  \includegraphics[width=0.86\textwidth]{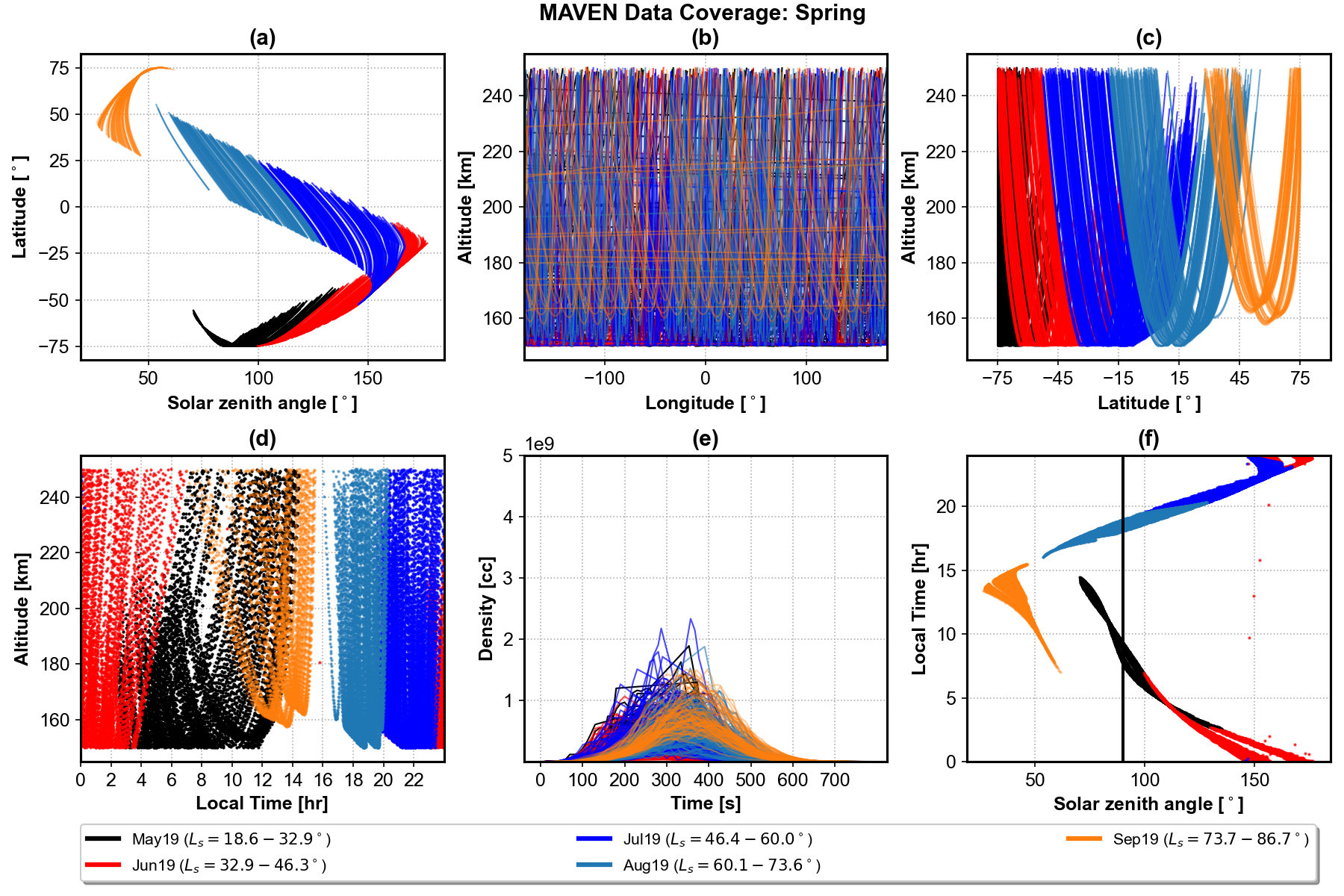}
  \hspace*{-1.cm} \centering
  \includegraphics[width=0.86\textwidth]{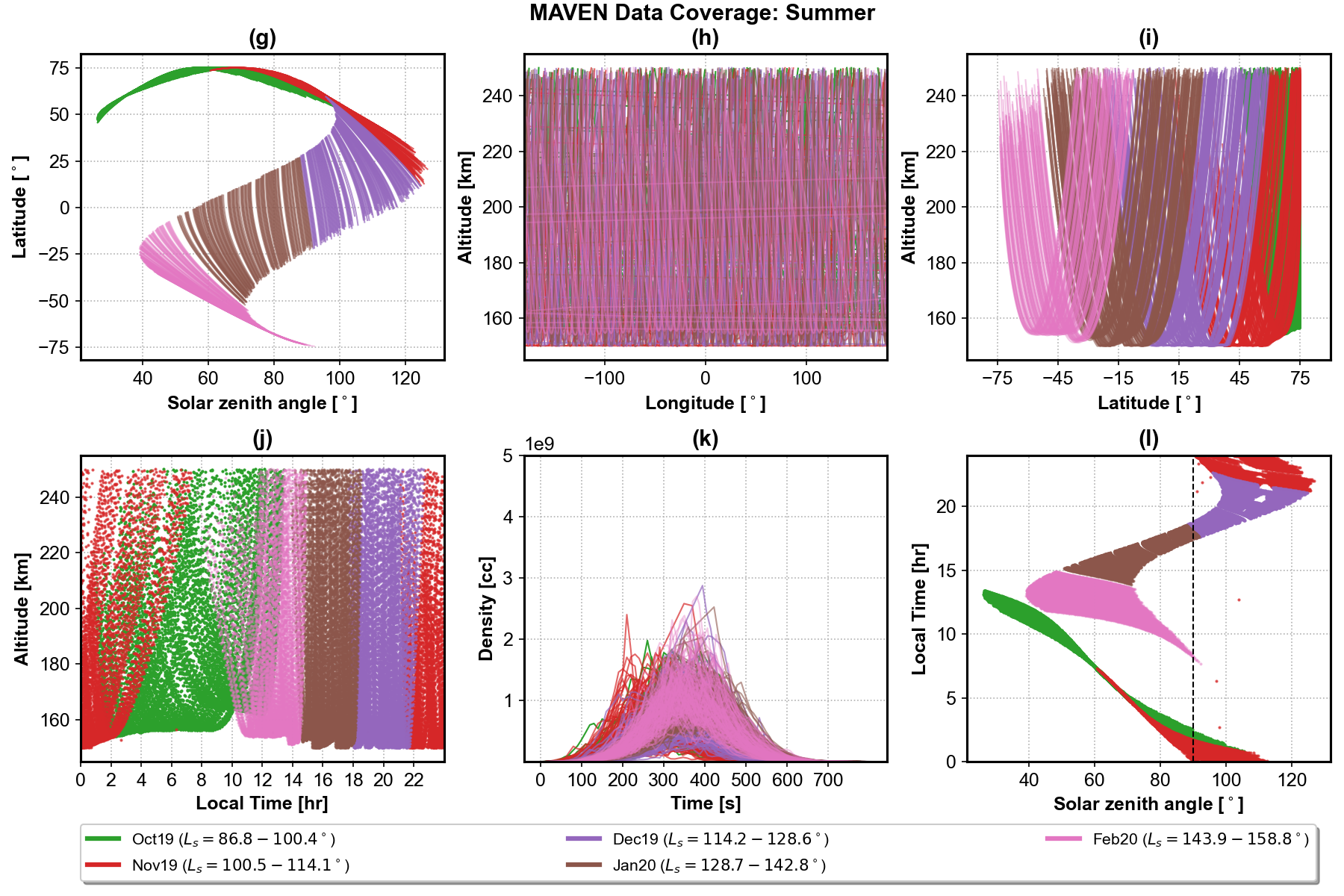}
  \caption{MAVEN data coverage during solar minimum in MY 35 with solar longitudes $L_s = 18.6-158.8^\circ$ corresponding to northern hemisphere spring (May 2019 to September 2019) and summer (October 2019-February 2020) conditions. The orbital coverage of the chosen spring-time data are shown in terms of (a) latitude-solar zenith angle, (b) altitude-longitude, (c) altitude-latitude, (d) altitude-local time, (e) density-time, and (f) local time-solar zenith angle distributions. The summer-time data coverage is shown in the same format in panels (g)-(l).}  \label{fig:solar_min_coverage}
\end{figure*}

In order to study the climatology of the Martian thermosphere and gravity wave activity at solar minimum, we analyzed carbon dioxide (CO$_2$)  number density data from the Neutral Gas and Ion Mass Spectrometer \citep[NGIMS,][]{Mahaffy_etal15} instrument on board NASA's Mars Atmosphere and Volatile Evolution  \citep[MAVEN,][]{Bougher_etal15b} spacecraft. The MAVEN/NGIMS has been performing systematic measurements in the Martian upper atmosphere since the beginning of its prime science mission in November 2014 and, thus, its observations cover Mars over the last solar minimum. MAVEN has an orbital period of about 4.5 hours, an inclination of 75$^\circ$, and a nominal periapsis of 150-160 km.

NGIMS is a quadrupole mass spectrometer with a mass range of 2-150 Da and a unit mass resolution. It was designed to fully characterize the abundances of 20 neutrals and ions in the Martian upper atmosphere. NGIMS collects its measurements every orbit when the MAVEN spacecraft descends below 500 km. The standard  error of individual measurements due to random uncertainties  depends on the ambient density and is typically $\sim$10\% at $ 7\times 10^5$ cm$^{-3}$, and is less than  $1\%$  when number density drops below $5\times 10^5$ cm$^{-3}$. 

We  analyze \textit{ten months} of density data around the solar minimum that took place in December 2019, i.e., 1 May 2019 to 29 February 2020, covering a period of solar longitudes $L_s=18.6^\circ- 158.8^\circ$ in Martian Year (MY) 35. This interval includes northern hemisphere spring and summer seasons and encompasses a total of 1307 MAVEN orbits with 618 northern spring and 689 summer season orbits, as summarized in Table \ref{tb:orbit_data}. In the rest of the study, we  consider the data for the months of May 2019-September 2019 ($L_s = 18.6-86.7^\circ$) and October 2019-February 2020 ($L_s = 86.8-158.8^\circ$)  to be representative of the northern spring and summer conditions, respectively. The associated spatiotemporal coverage  by the spacecraft during the chosen period is presented in Figure~\ref{fig:solar_min_coverage} for spring (a-f) and summer (g-l) periods separately, with data organized according to the different months. While NGIMS scans altitudes  up to 500 km, the signal above $\sim$ 250 km is too weak, and is not suitable for this study. If used, these densities would affect the polynomial fit of the background density and, consequently, the wave-induced density fluctuations described in Section~\ref{sec:retr-grav-wave}. 

The spring data covers solar zenith angles of $\chi = 20^\circ-170^\circ$, which are shown as a function of latitude (Figure~\ref{fig:solar_min_coverage}a). Nighttime as well as daytime measurements are available, however the latitude varies significantly as a function of the solar zenith angle. The longitude and latitude ($75^\circ$S-$75^\circ N$) coverage is nearly global (panels b-c), with some gaps in the northern hemisphere midlatitudes, and the different local times are well represented, except  in the 15-17 h interval during the spring season (panel d). The peak CO$_2$ densities are up to $\sim 2\times 10^9$ cm$^{-3}$. They are plotted as a function of the duration of the orbit (inbound plus outbound passes), which is typically 700 s (panel e). Variability of the orbit during the chosen period is also highlighted by changing correlation between the solar local time and solar zenith angle due to slight seasonal changes (panel f). The solar zenith angle $\chi = 90^\circ$ is marked in the latter panel, as it is used  for separating the daytime and nighttime data in the analysis to be presented. Orbital coverage during the chosen summer season (Figure~\ref{fig:solar_min_coverage}g-l) is overall similar to the spring season, however, the range of solar zenith angles is slightly narrower: $\chi = 20^\circ-120^\circ$, while the latitude and local time coverages are slightly better than during the spring season. Similar orbital coverage during spring and summer allow for an intercomparison of GW activity between  these seasons, which  are studied in section~\ref{sec:global-local-time}.

\begin{figure}[h!]
        \centering
        \includegraphics[width=\textwidth]{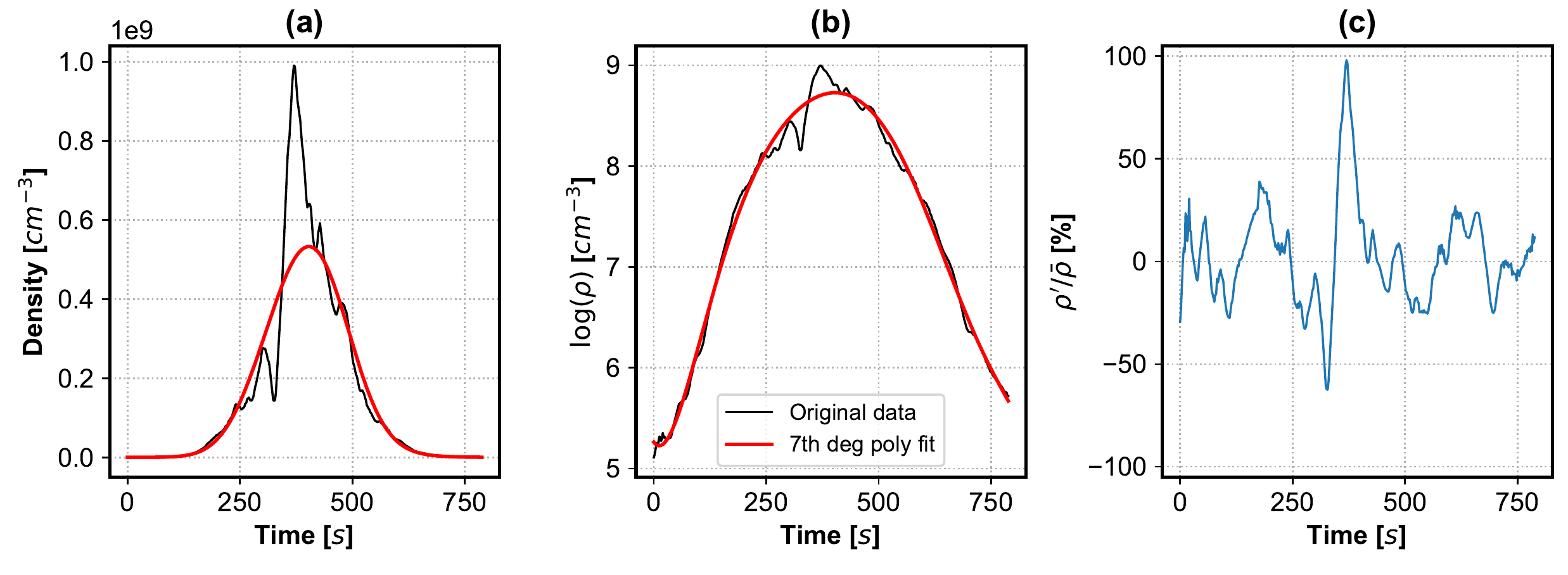}
        \caption{Illustration of the retrieval of GW activity from a single profile on December 1st, 2019, 02:40:06 UT. Variations of (a)  CO$_2$ density in cm$^{-3}$ and the 7-th order polynomial fit, (b) logarithm of the CO$_2$ density and 7-th order fit, (c) relative GW-induced density perturbations (in percentage) during one orbit as a function of time (in seconds). Black line shows the instantaneous density measurement along the orbit. Red line shows the fitting of the mean values using a 7-th order polynomial. }
        \label{fig:sing_orbit_gw_activity}
\end{figure}

\subsection{Retrieval Technique for Gravity Wave Activity}
\label{sec:retr-grav-wave}
Effects of gravity waves can be identified in terms of fluctuations in thermodynamic parameters, such as temperature, density, and winds, around the respective mean value associated with the large-scale global circulation and atmospheric features. MAVEN  measures the instantaneous thermospheric density, which can be decomposed into a background (mean) and a deviation components:
\begin{equation}  \label{eq:density}
\rho = \rho^\prime + \bar{\rho},
\end{equation}
where $\bar{\rho}$ is the background density, the bar denotes an appropriate spatiotemporal average of the density (the so-called a Reynolds average), and $\rho^\prime$ is the density wave disturbances. Here we specifically concentrate on carbon dioxide density measurements. If the background density is known, then the fluctuating component, which is a proxy for GW activity, can be quantified. For this, we first determine the background component by fitting a 7-th order polynomial to the logarithm of the CO$_2$ density, as has been done in earlier studies \citep[e.g.,][]{Jesch_etal19,Yigit_etal15b, Yigit_etal21b}. The procedure is illustrated in Figure~\ref{fig:sing_orbit_gw_activity}, where the variations of the instantaneous density and the 7-th order polynomial fit (i.e., mean density) along with the resulting GW perturbations in percentage are shown as functions of orbit duration. Both inbound and outbound passes are used for a given orbit in our analysis. 

\subsection{Data Binning, Monthly and Daily Mean Gravity Wave Activity}\label{sec:data-bins}

In order to study the spatiotemporal variations of  GW activity, we bin the relative density fluctuations $\rho^\prime/\bar{\rho}$ calculated for each orbit (inbound and outbound passes) in terms of altitude, latitude, local time, and solar zenith angle with bin sizes of 10 km, 5$^\circ$, 1 h, and 10$^\circ$, respectively. Performing this procedure for a given month yields the monthly mean GW activity $\overline{\rho^\prime/\bar{\rho}}$ to be presented in Section~\ref{sec:grav-wave-activ}. In the calculation of the daily mean GW activity, the data is  grouped in 10 km altitude bins. In the averages of the GW activity to be presented, the typical fractional uncertainty is less than 1\%. Further details of the error analysis can be found in Appendix~\ref{sec:uncert-grav-wave}.


\section{Results}\label{sec:results}
\subsection{Variations of Gravity Wave Activity during Solar Minimum}\label{sec:grav-wave-activ}
\begin{figure*}[t!]
  \vspace{-0.1cm}
   \includegraphics[width=1.0\textwidth]{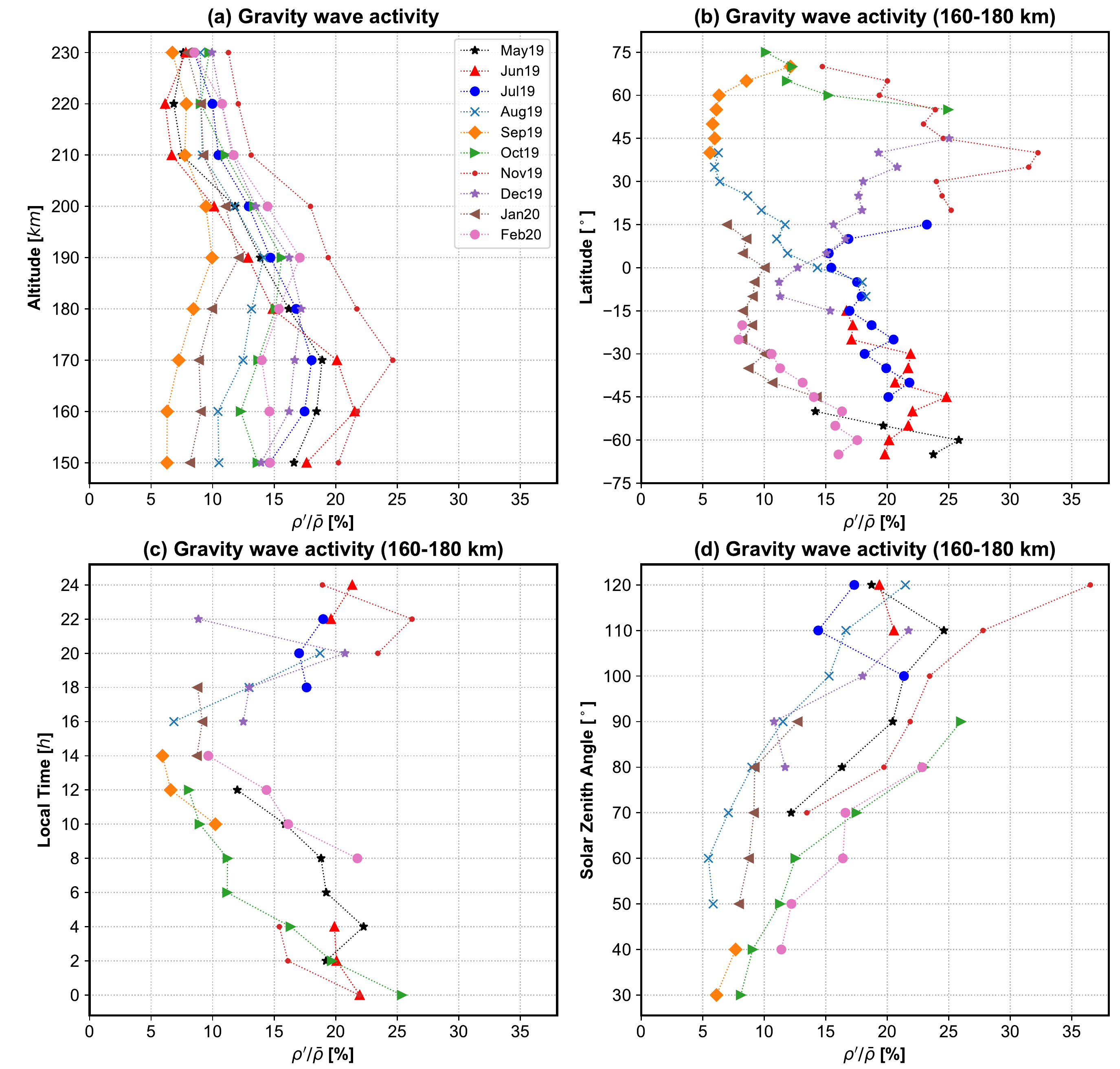} 
  \caption{Altitude, latitude, local time and solar zenith angle variations of the monthly mean gravity wave activity in terms of relative density fluctuations from May 2019-February 2020, corresponding to $L_s =  18.6^\circ - 158.8^\circ$, representative of northern spring and summer seasons.}
  \label{fig:GW_activity_solar_min}
\end{figure*}

We next present the climatology of GW activity around the solar minimum. In order to provide a better global picture, we chose ten months of data (May 2019 - Feb 2020, $L_s = 18.6-158.8^\circ$), representative of spring and summer seasons, as shown in Figures~\ref{fig:solar_min_coverage}. Figure~\ref{fig:GW_activity_solar_min} shows the distribution of wave-induced relative density fluctuations. The results are presented as monthly averages binned with respect to altitude, latitude, local time, and solar zenith angle shown in panels (a)-(d), respectively. The month-to-month variations of GW activity are noticeable, which are due largely to a combination of seasonal variations in GW propagation and dissipation and the changes in MAVEN's orbital coverage. Overall, the GW activity varies between $\sim5-35\%$ during this period. The smallest observed magnitudes of disturbances are found in September 2019 ($L_s = 73.7-86.7^\circ$) with $5-10\%$ around the local noon at middle-to-high latitudes of the northern hemisphere. GW activity maximizes in November 2019 ($L_s = 100.5-114.1^\circ$) between 160-190 km at northern hemisphere middle-latitudes at nighttime. Above 180-190 km, amplitudes of GWs  decay with altitude for all months. Yet, around 230 km there is still appreciable wave activity of $5-15\%$.

The latitude, local time, and solar zenith dependencies are given for the altitude region of 160-180 km, where GW activity typically peaks. In general, $\overline{\rho^\prime/\bar{\rho}}$ exhibits strong local time variations with larger nighttime  values than daytime. It is noticeable that the GW activity increases with solar zenith angle during the analyzed period, in accordance with the local time dependence. The large difference in GW activity seen between the northern and southern hemispheres is primarily due to a summer-to-winter change of GW activity amplified by the variations of the local time as observed by the spacecraft during this period. Longitudinal variability of GW activity, although non-negligible, has been found to be weaker than other variations (and, therefore, not shown here), which suggests that the thermospheric GW activity is influenced by topographical features to a lesser extent.  

\begin{figure*}[t!]
  \hspace*{-0.cm}
  \centering\includegraphics[width=0.8\textwidth]{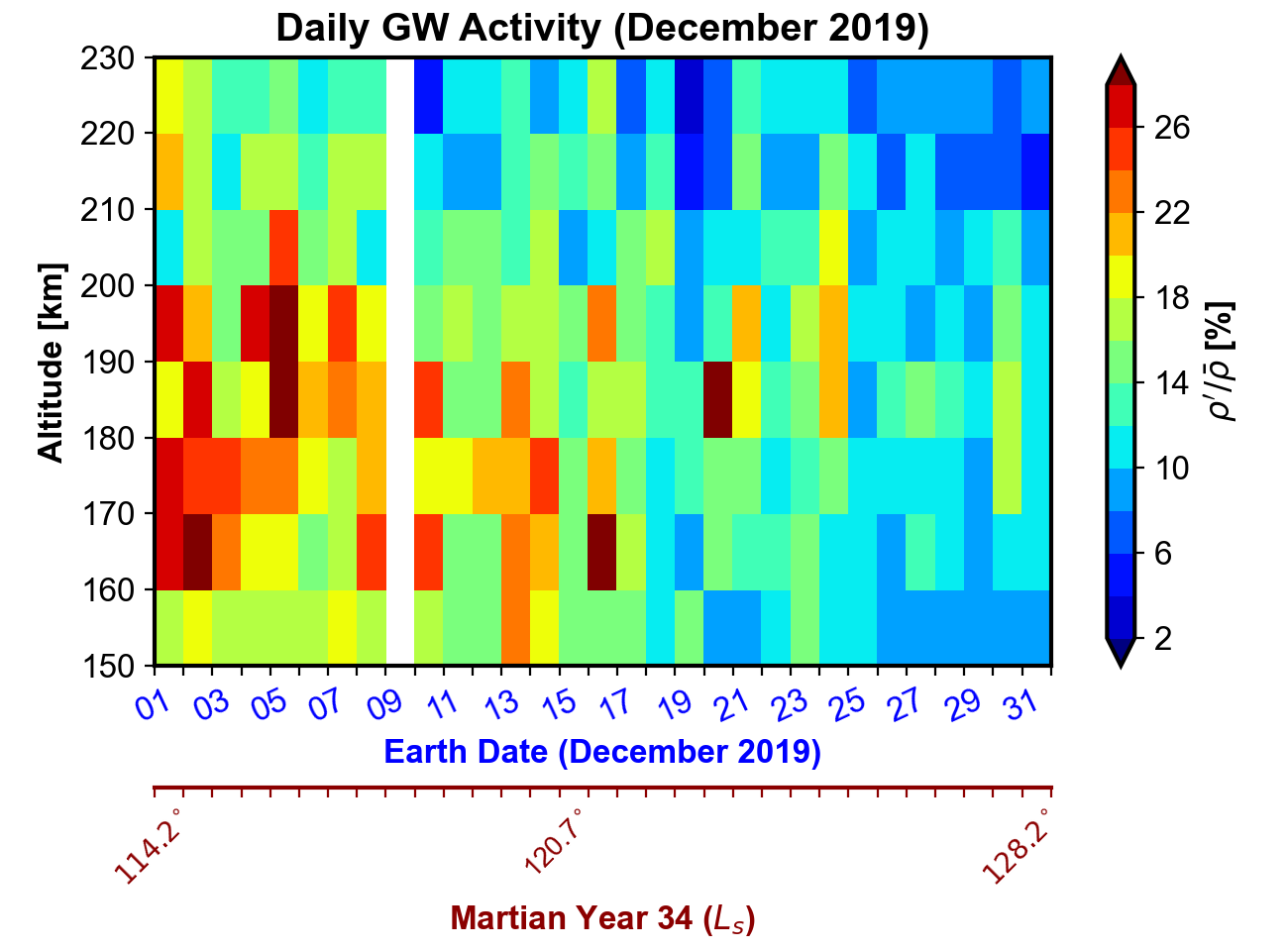}
  \caption{Daily mean GW activity during December 2019 in terms of relative density fluctuations. Earth date and solar longitude are shown in the x-axis. White space in the plots denotes data that are either missing or not suitable for scientific use.}  \label{fig:GW_activity_daily_dec19}
\end{figure*}

\begin{figure*}[t!]
  \hspace*{-0.cm}
  \centering\includegraphics[width=0.9\textwidth]{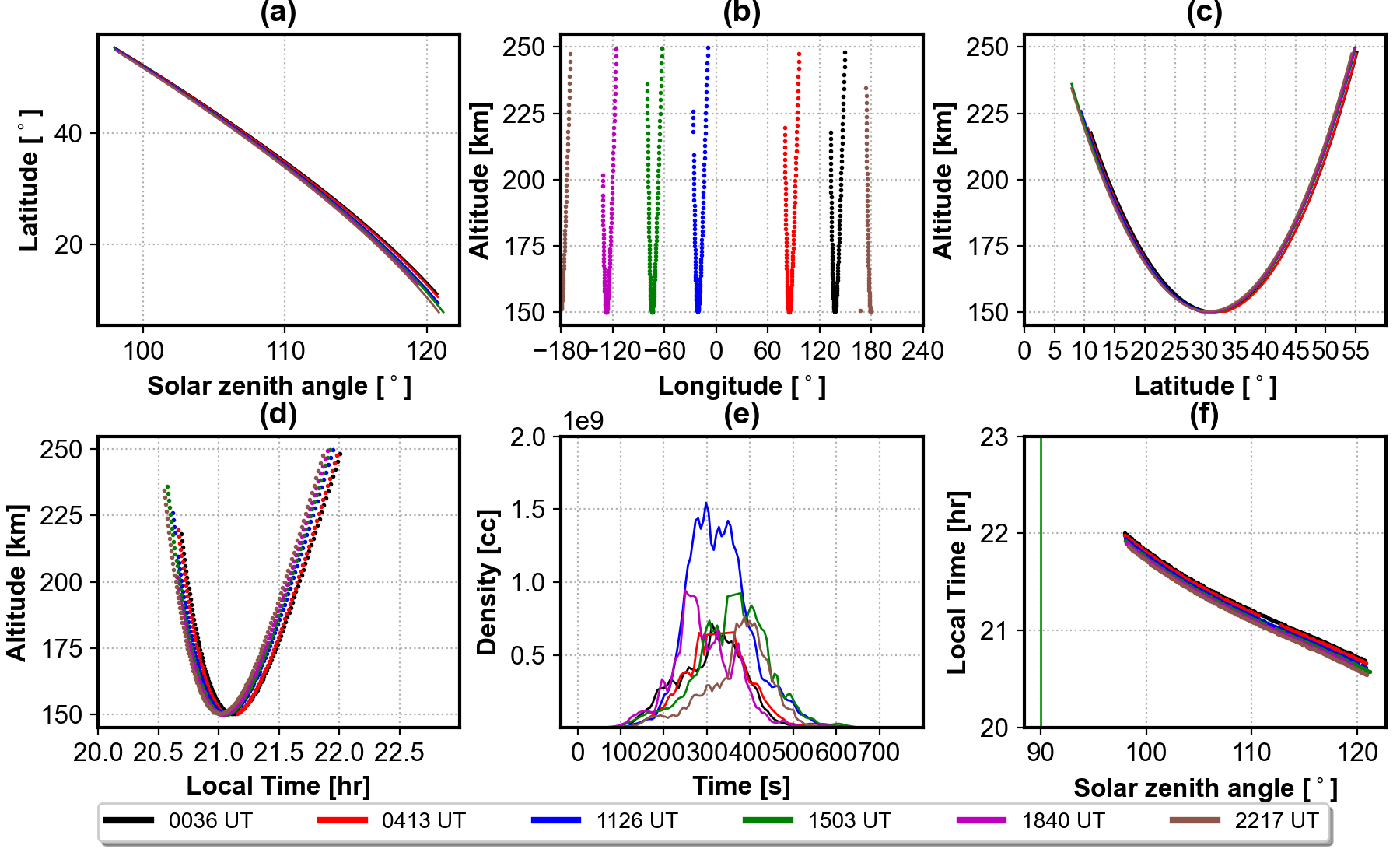}
  \caption{MAVEN's orbital coverage on 5 December 2019. The panel lists the specific orbits included in the plot with respect to universal times (UT).}
  \label{fig:cov_5dec19}
\end{figure*}

Gravity wave propagation and dissipation are highly variable processes due to changes in wave sources and the background atmospheric fields. One way of illustrating GW variability is to study the day-to-day variation of the relative density fluctuations. For this, we singled out the solar minimum month, December 2019, binned the individual orbits daily and calculated the daily mean GW activity, as shown in the altitude versus day number distribution in Figure~\ref{fig:GW_activity_daily_dec19}. During December 2019, Mars was close to aphelion ($L_s=114.2^\circ-128.6^\circ$), and the seasonal changes were relatively small. The MAVEN spacecraft was completely on night side, moving southward and towards the day side (Figure~\ref{fig:solar_min_coverage}). The respective coverage was 55$^\circ$N--20$^\circ$S for latitude, 19--22 h for local time, and 80$^\circ$--110$^\circ$ for the solar zenith angle. The late December coverage corresponds to slightly earlier local times (still night-time ones) and is closer to the equator than in early December. Overall, the day-to-day variations of the wave activity are appreciable, despite the relatively small day-to-day changes in the orbital configuration. Magnitudes of relative density disturbances in the early December (up to 30\%) are significantly larger than those in late December, when they drop to $\sim $10\%. The peak activity is often found between 160-200 km. 

In December 2019, the largest GW activity is seen on 5 December 2019 (Figure \ref{fig:GW_activity_daily_dec19}), which we study next in more detail by zooming in into the orbit-to-orbit variations of the relative density fluctuations. The corresponding orbital coverage is shown in Figure~\ref{fig:cov_5dec19}. Within 5 December 2019, MAVEN has completed 6 orbits, which are represented by different colors corresponding to different UTs. Except for the longitude variations, all other geophysical parameters vary to a minor degree from orbit to orbit. Overall, the latitudes of  $5-55^\circ$N, local times of 20.5-22 h with solar zenith angles of 100-120$^\circ$ are observed. The peak density of $\sim 1.5\times 10^9$ cm$^{-3} $ is found at 1126 UT at periapsis around $-20^\circ$ longitude and $30^\circ$N latitude. 

\begin{figure*}[t]
  \hspace*{-0.cm}
    \centering\includegraphics[width=0.9\textwidth]{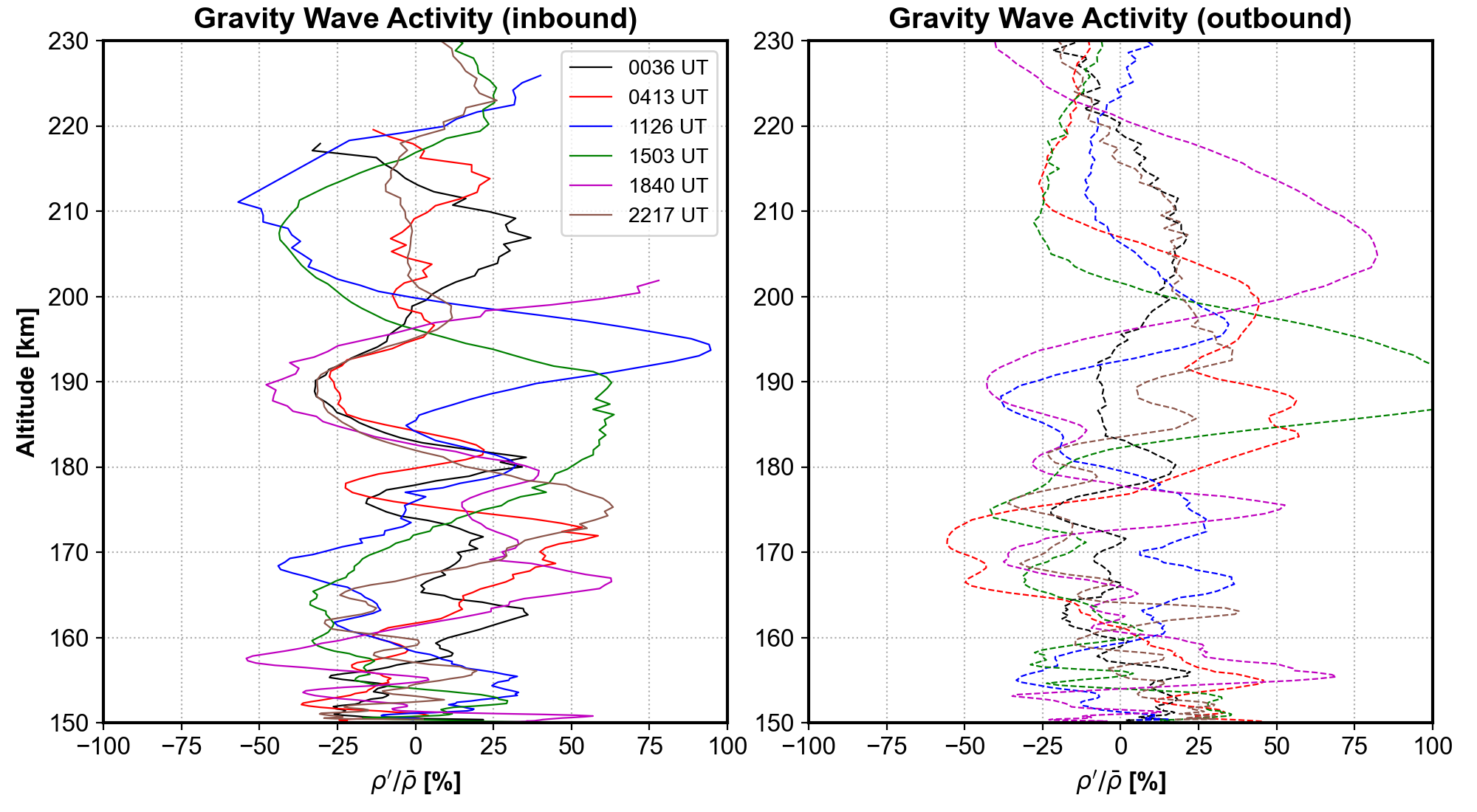}
  \caption{Altitude variations of gravity wave activity in terms of relative density fluctuations during the different (a) inbound and (b) outbound passes within the six orbits on 5 December 2019. The different orbits are represented by different colors in terms of their associated universal times.}
  \label{fig:gw_5dec19}
\end{figure*}

The altitude variations of the GW activity during the six orbits on 5 December are plotted in Figure~\ref{fig:gw_5dec19}, presented by subdividing each orbit into their inbound and outbound passes shown in panels (a) and (b), respectively. The instantaneous values of GW activity vary significantly (up to $\pm 50\% $), occasionally jumping up to $100\%$. Orbit-to-orbit variations are noticeable, which points out to a longitudinal variability of GW activity. There are major differences between the inbound and outbound passes for a given orbit as well, which is indicative of latitudinal variability, since the inbound and outbound passes correspond to somewhat different latitudes, in this case varying between $5-30^\circ$N and $30-55^\circ$N, respectively.

\subsection{Global and Local Time Distribution of Gravity Wave Activity}\label{sec:global-local-time}

\begin{figure*}[t!]
  \hspace*{-0.5cm}
  \centering\includegraphics[width=0.95\textwidth]{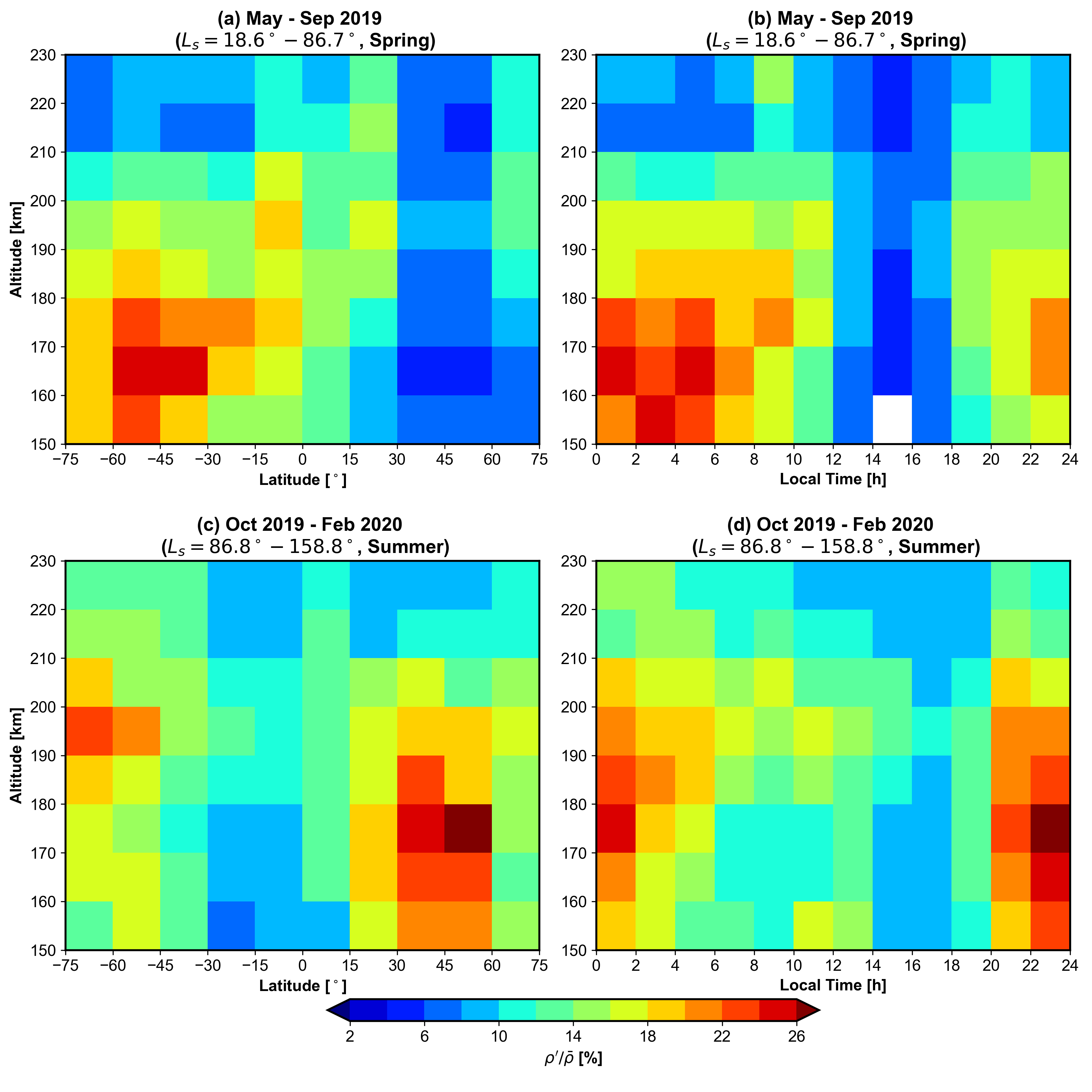}
  \caption{Altitude versus latitude and local time distributions of gravity wave activity in northern hemisphere spring (a-b) and northern hemisphere winter (c-d) during solar minimum period (May 2019 to February 2020 $L_s=18.6^\circ - 158.8^\circ$) as observed by MAVEN/NGIMS. The data is binned in terms of 10 km $\times$ 15$^\circ$ altitude $\times $latitude bins. The same color scale is used for all figures.}
  \label{fig:GW_activity_solar_min_ALT}
\end{figure*}

\begin{figure*}[t!]
  \hspace*{-0.5cm}
  \centering\includegraphics[width=1.0\textwidth]{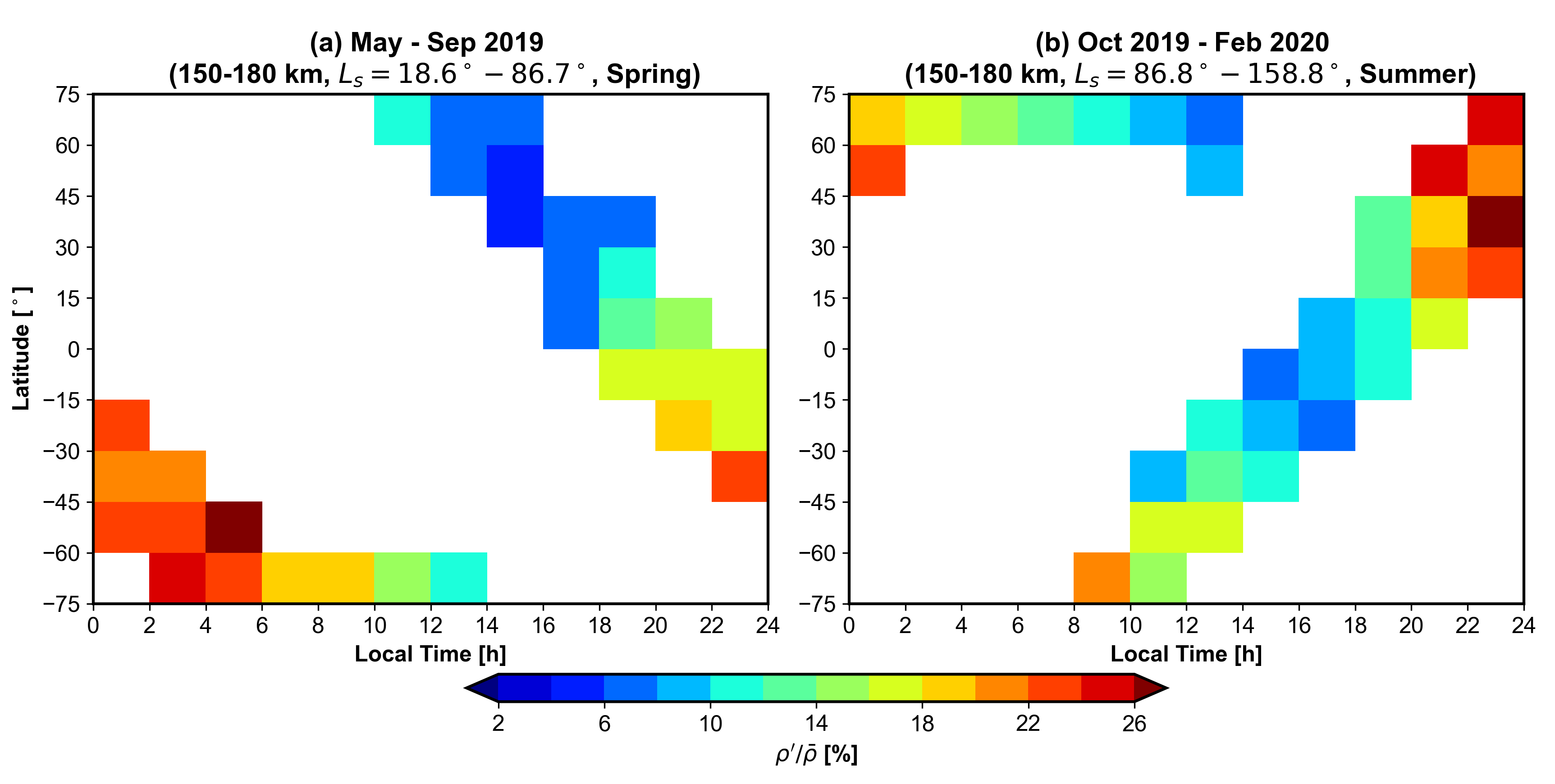}
  \caption{Latitude-local time distributions of gravity wave activity between 150-180 km during (a) northern spring and (b) northern summer as observed by MAVEN/NGIMS. The data is binned in terms of 15$^\circ \times$ 2-hour  latitude$\times $local time bins. The same color scale is used for all figures.}
  \label{fig:GW_activity_Lat_LT}
\end{figure*}

Figure~\ref{fig:GW_activity_solar_min_ALT} shows the altitude variations between 150 and 230 km of the relative density fluctuations during northern spring (upper panels) and summer (lower panels), binned as a function of latitude (left panels) and local time (right panels). Each seasonal result is based on  a five-month average, considering all NGIMS data from August 2019 to September 2019 ($L_s=18.6^\circ - 86.7^\circ$) and from October 2019 to February 2020 ($L_s=86.8^\circ - 158.8^\circ$) as representative of northern hemisphere spring and summer, respectively. MAVEN orbital coverage is  similar globally for the chosen seasons, as shown above in Figure~\ref{fig:solar_min_coverage}, with a good local time and latitude ($75^\circ$S-$75^\circ$N) coverage.

During northern spring season, GW activity maximizes with $\sim 26\%$ in the southern hemisphere high-latitudes around 160-170 km. Generally, GW-induced fluctuations of density are much larger in the southern hemisphere, but this difference should be interpreted with caution since the northern hemisphere midlatitudes are poorly sampled by MAVEN during spring. The nighttime GW activity is much larger than the daytime values: the magnitudes peak around 0-6 h between 150-180 km with $\sim 26\%$ and can be as low as a few percent around the local noon. During northern summer season, larger GW activity is found in the northern hemisphere midlatitudes around 170-180 km. Larger nighttime GW activity is noticeable, but the day-night difference is slightly weaker than during northern spring season. The peak of GW disturbances occurs around the local midnight. The spring and summer seasons reveal different pictures of GW activity. However, the most remarkable feature that transpires from the data is the persistent day-night contrast of GW activity. It is also seen that GW fluctuations are larger in the upper thermosphere (above 200 km) during northern summer compared to those in spring, which indicates that  waves experience more favorable conditions for propagating upward during northern summers.

It is seen from the orbital coverage data (Figure~\ref{fig:solar_min_coverage}) that the local time changes when the spacecraft moves in the latitude direction.  Therefore, in order to facilitate a more consistent analysis of seasonal differences, we next present in Figure~\ref{fig:GW_activity_Lat_LT} the latitude-local time cross sections of the GW activity in summer and spring. It shows that the lowest GW activity in the northern hemisphere midlatitudes occurs around the local noon, while the high-latitude GW peak coincides with nigthttime. Similarly, during northern summer, the midlatitude GW maximum occurs around local midnight. This analysis suggests that, due to the spacecraft orbital variation in latitude and local time, it can be challenging to consistently compare the different seasons. Nevertheless, it is clearly seen that GW activity maximizes at nighttime regardless of latitude.

\begin{figure*}[t!]
  \hspace*{-0.5cm}
  \centering\includegraphics[width=1.0\textwidth]{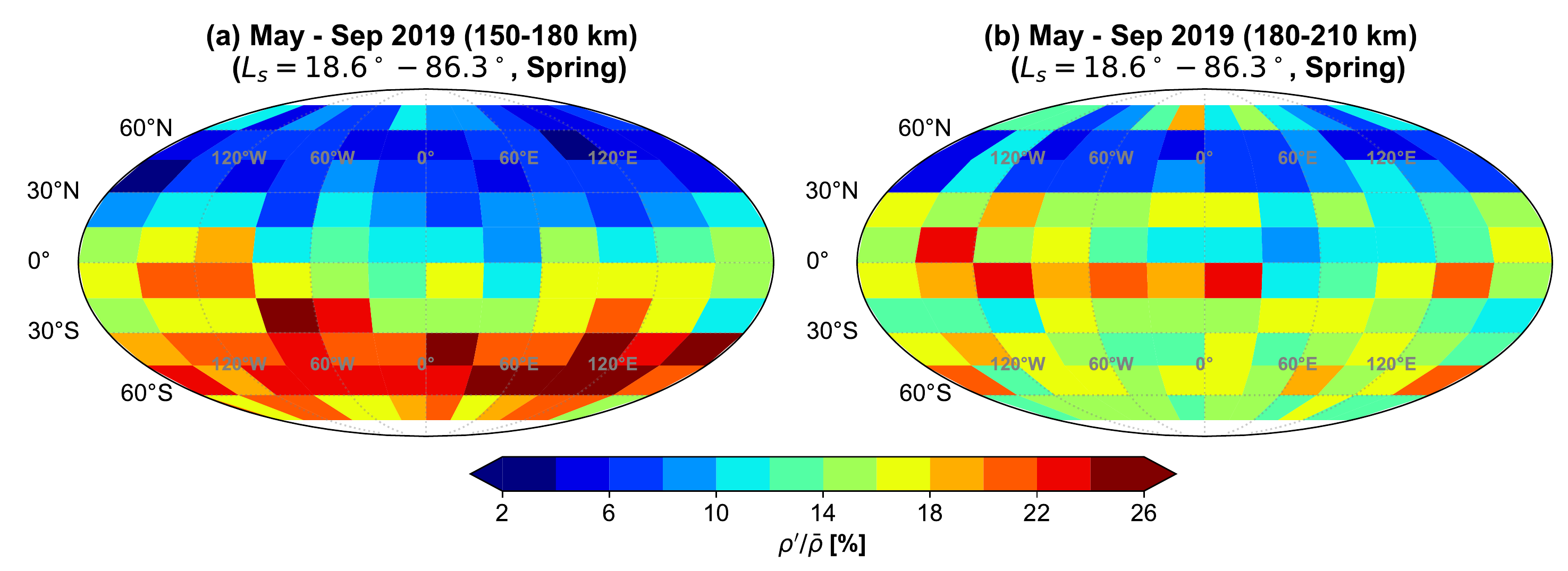}
  \hspace*{-0.5cm}
    \centering\includegraphics[width=1.0\textwidth]{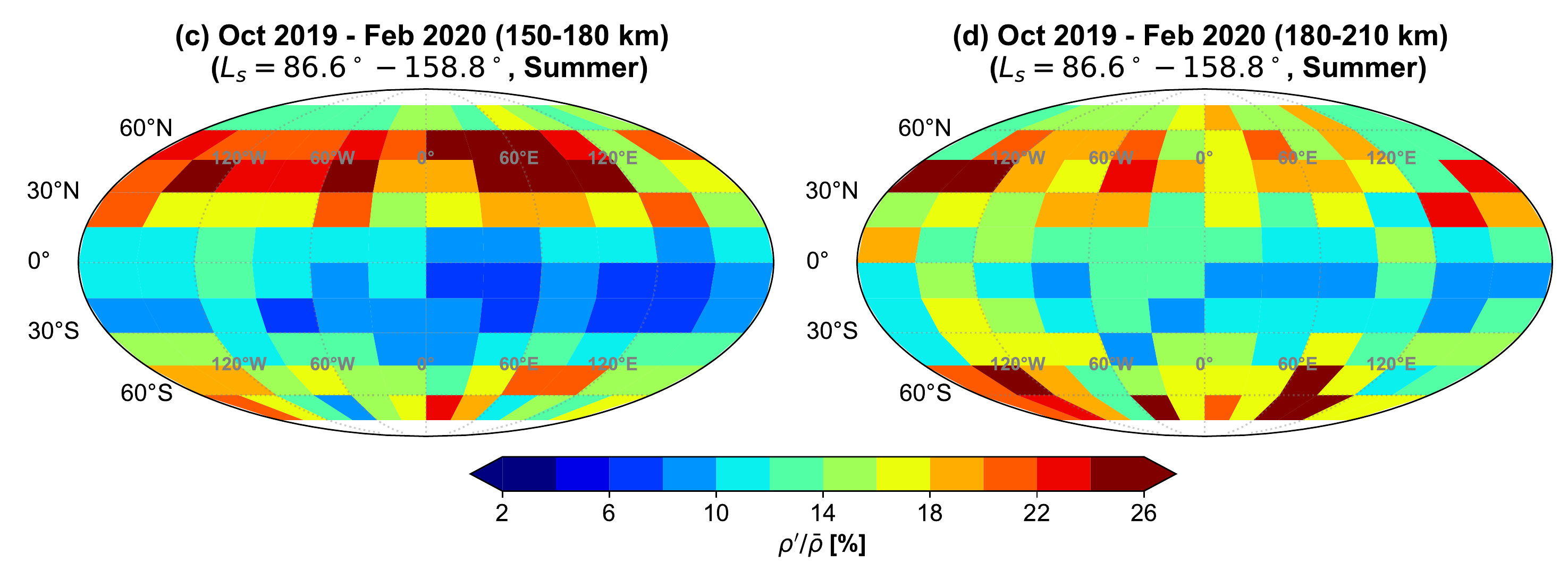}
  \caption{Global distributions of gravity wave activity around solar minimum period from
in northern hemisphere spring (a-b) and northern hemisphere winter (c-d) for the altitude range 150-180 km (a,c) and 180-210 km (b,d) during the chosen solar minimum period (May 2019 to February 2020 $L_s=18.6^\circ - 158.8^\circ$) as observed by MAVEN/NGIMS, as observed by MAVEN/NGIMS. The data is binned in terms of $15^\circ \times 30^\circ$ latitude $\times$ longitude bins. The same color scale is used for all figures.}
    \label{fig:GW_activity_solar_min_LATLON}
\end{figure*}

Figure~\ref{fig:GW_activity_solar_min_LATLON} presents the global distributions of GW activity  at two representative altitudes: averaged between (a) 150-180 km and (b) 180-210 km. Regions poleward of 30$^\circ$ clearly exhibit larger GW amplitudes compared to that in low-latitudes. Thus, magnitudes of relative density fluctuations at 150-180 km are up to $6-10\%$ equatorward of 30$^\circ$, while they increase up to 20\% in the middle- to high-latitudes. A similar pattern is seen at 180-210 km with larger high-latitude values of $\sim$26\%. While the distributions of the observed relative density fluctuations suggest a clear hemispheric asymmetry at both representative altitude regions, the degree of the actual asymmetry is difficult to assess without a dedicated three-dimensional modeling study, since the local time coverage (as well as the solar zenith angle) during the observed period changes with the orbit. The longitudinal variability of the GW activity is less apparent, as was mentioned above.


\section{Discussion}\label{sec:discussion}
\subsection{Gravity Wave Activity around the Solar Minimum}\label{sec:grav-wave-activ-1}

The Martian thermosphere is the uppermost region of the atmosphere, where spacecraft perform aerobraking maneuvers and atmospheric species can escape to space. Characterization of  GWs in the upper atmosphere is essential for understanding its structure and dynamics. In this study, we have characterized the thermospheric GW activity above 150 km at around the recent solar minimum corresponding to the end of Solar Cycle 24, as observed by NASA's Mars Atmosphere and Volatile Evolution (MAVEN) \citep{Bougher_etal15b} spacecraft. The Martian weather is extremely variable and produces a broad spectrum of internal waves \citep{Ando_etal12}. GW packets propagate to higher altitudes, where they can be detected by satellites. A number of previous studies have identified the presence of GWs at thermospheric altitudes using Mars Global Surveyor (MGS) and Mars Odyssey measurements \citep{Creasey_etal06b, Fritts_etal06} and MAVEN data \citep[e.g.,][]{Yigit_etal15b, England_etal17, Siddle_etal19, Li_etal21}. MGS accelerometer data covered the thermosphere from 100 to 170 km, while MAVEN's typical periapsis altitude is around 150 km, thus MAVEN covers middle and upper regions of the thermosphere. Numerical modeling efforts suggested a dynamical significance of GWs in the Martian atmosphere and highlighted their essential role in coupling the lower and upper atmospheric layers \citep{Parish_etal09, Medvedev_etal15, Yigit_etal18, Kuroda_etal19}. During a minimum of solar activity, GW disturbances of lower atmospheric origin become more observable in the thermosphere, because the upper atmosphere is less disturbed by magnetic and solar process. Our study reveals systematic spatiotemporal distributions of thermospheric GWs with relative density fluctuations ($\rho^\prime/\bar{\rho}$) reaching 30-35\% around 160-180 km. Instantaneous magnitudes of GW activity can exceed 100\%. The month-to-month variations show a significant degree of variability due to a combination of changes in the spacecraft orbital parameters and physical processes controlling GW propagation and dissipation. Note that it is often challenging to decouple local time and latitude variation due to MAVEN's orbital precession.

Overall, the vast majority of the previous MAVEN studies of GWs focused on observations during moderate solar activity conditions. While there are differences in the individual values of GW activity presented by these studies due to the orbital coverage being used, in general, the range of wave magnitudes presented here is similar to previous observations. For example, the study by \cite{Leelavathi_etal20}  based on the MAVEN data for MY33 and 34 reported the mean thermospheric GW amplitudes up to 30\%. \cite{Terada_etal17}  analyzed MAVEN data in the early phase of the mission from February 2015 to March 2016 and found root-mean-square GW amplitudes of $6-30\%$.

The majority of our GW study is based on a statistical analysis, as required by the nature of gravity wave process, by binning data in vertical, local time, latitude and longitude intervals (e.g., Figures 4, 7, 8, 9). Within each bin, GWs are averaged. The horizontal distance covered near the periapsis depends on latitude and longitude. At altitudes between around 150-155 km, where the spacecraft travels predominantly horizontally, measurements pick up contributions from waves propagating mainly horizontally.
We next consider processes that affect the observed thermospheric GW activity. 

The amount of GWs in the thermosphere is determined by the nature of the wave sources in the lower atmosphere, propagation conditions and damping. On Mars, the main mechanisms of GW generation are flow over topography (orographic sources), instability of weather systems, and convection \citep{Heavens_etal20a}. Observations \citep{Creasey_etal06a} and simulations with high-resolution general circulation models \citep{Kuroda_etal19} show that GW sources are highly intermittent and vary spatially with seasons. In average, the nonorographic generation dominates in the lower atmosphere, while magnitudes of wave disturbances remain relatively constant.  The exception is the periods of major dust storms, when GW activity drops in the lower atmosphere \citep{Heavens_etal20a, Kuroda_etal20}. The amplitudes of GWs below $\sim$30 km are generally larger in low latitudes, unlike in the upper thermosphere, as reported above. Obviously, GW sources alone cannot explain the observed distribution of GWs in the thermosphere.

Vertical propagation of GWs can be considered using detailed numerical wave models \citep[e.g.,][]{Parish_etal09, Heale_etal14, KshevetskiiGavrilov05}. However, as far as only mean quantities are concerned, it is more straightforward to assess the evolution of averaged (devoid of phase information) wave characteristics with height. In particular, the vertical flux of the horizontal wave momentum ($\bar{\rho}\overline{u^\prime w^\prime}$) is a suitable quantity, because it is conserved, if no dissipation is applied. The appropriate equation can be obtained by applying vertical differentiation to Eqn. 1 in the paper of \citet{Yigit_etal08} taking account of their Eqn. 2:
\begin{equation}
   \frac{d\overline{u^\prime w^\prime}}{dz} = \Bigl(-\frac{\bar{\rho}_z}{\bar{\rho}} - \beta_{mol} 
    - \beta_{non} \Bigr)\overline{u^\prime w^\prime}.
    \label{eq:flux}
\end{equation}  
In (\ref{eq:flux}), $u^\prime$ and $w^\prime$ are fluctuating (wave) components of the horizontal and vertical wind, correspondingly; the overline denotes averaging over wave phases, $\bar{\rho}$ is the background density; $\beta_{mol}$ and $\beta_{non}$ are the vertical damping rates due to molecular diffusion and nonlinear breaking/saturation processes, respectively; $\bar{\rho}_z/\bar{\rho} =-1/H$ with $H$ being the density scale height.
Without dissipation ($\beta_{mol}=\beta_{non}=0$), the flux per unit mass $F=\overline{u^\prime w^\prime}$ grows exponentially with height due to decreasing background density. While GW amplitudes are relatively small near the sources, they can reach substantial values at thermospheric heights, as  is clearly demonstrated in altitude distributions of wave activity (Figures \ref{fig:GW_activity_solar_min}a, \ref{fig:GW_activity_daily_dec19}, and \ref{fig:GW_activity_solar_min_ALT}a). Note that $F$ is a quadratic quantity of amplitude, therefore the wave amplitude itself is proportional to the square root of $F$.  

The decay of GW activity above 180-190 km is an indication of increased wave dissipation.  In the middle atmosphere, nonlinear breaking/saturation $\beta_{non}$ strongly limits, and even completely erases, GW harmonics with small vertical wavelengths. 

According to the classical linear convective instability theory,  a monochromatic GW becomes convectively unstable, when the wave perturbation causes the total lapse rate to become superadiabatic.  This is described mathematically by
      \begin{equation}
        u^\prime \ge |c_i-\bar{u}|,  
      \end{equation}
      where $c_i-\bar{u}$ is the intrinsic horizontal phase speed of the probe wave ``i".
    
Nonlinear diffusion is a mechanism that extends the concept of linear instability to a broad spectra of nonlinearly interacting GWs \citep{MedvedevKlaassen95}. For a single self-interacting harmonic, it yields the convective instability threshold of \cite{Lindzen81}, except for the damping occurring not suddenly, but starting at some levels below and reaching the maximum at the linear threshold \cite[Sect. 7]{MedvedevKlaassen00}.  According to the theory, the damping acting on a wave harmonic ``$i$" due to self-interactions as well as due to the effects of other waves is given by
\cite[Eqn. 9]{Yigit_etal08}
\begin{equation}
\beta_{non} = \frac{\sqrt{2\pi}N}{\sigma_i} \exp \Bigg(-\frac{|\hat{c_i}|^2}{2\sigma_i^2}\Bigg),
\end{equation}
where $N^2 = (g/T) \, (\partial T/\partial z + g/c_p)$ is the buoyancy frequency, with gravity $g$ and specific heat at constant pressure $c_p$, $\sigma_i$ is the wind variance produced by the waves with vertical scales shorter (i.e.,  intrinsic horizontal phase speed $\hat{c}_i = c_i - \bar{u}$ smaller) than the reference wave harmonic, i.e., $\sigma_i^2 = \sum_{m_j > m_i} \overline{u_j^{\prime 2}}$, $m=2\pi/\lambda_z$ being the vertical wave number.
The background wind enters the expression for $\beta_{non}$ and, thus, filters out harmonics moving in the same direction as the horizontal wind.

The formalism of nonlinear diffusion resolves some issues with the linear convective instability models of GWs. As was mentioned above, the linear instability criterion of  $u^\prime \ge |c_i-\bar{u}| $  implies a single breaking level, above which the GW ceases to exist. In the nonlinear theory, however, GWs continuously experience  damping at multiple levels, as their amplitudes grow. The GW drag (i.e., acceleration/deceleration produced by wave) estimated by the linear theory is often too large and requires tuning parameters (like ``intermittency" factors) to reduce them, while the nonlinear approach does not need fudge factors to adjust the GW forcing. Since GWs interact with the mean flow at multiple levels as accounted for in the nonlinear approach, the forcing is also distributed continuously over heights, while in the linear convective instability approach, GW drag is zero, if $u^\prime < |c_i-\bar{u}|$.

Mainly fast and large vertical-scale waves survive breaking upon propagation from below. 
For such waves that reach the thermosphere, the primary mechanism of wave  damping is molecular diffusion and thermal conduction ($\beta_{mol}$),  while the nonlinear wave damping $\beta_{non}$ remains secondary, as will be shown in the next subsection. Convective instability is less likely to occur at high altitudes in the thermosphere,  at least for waves originated in the lower atmosphere.  However, convective instability/ nonlinear damping can play an important role in the middle atmosphere in limiting wave amplitude growth.

The corresponding damping rate for the molecular diffusion and thermal conduction acting on a particular $i$-th GW harmonic from the spectrum \citep[Eqn. 14]{Yigit_etal08}, given by 
\begin{equation}
  \label{eq:molecular_dissipation}
  \beta_{mol}^i = \frac{\mu_{mol}\, N^3}{\bar{\rho}\,k_h (c_i-\bar{u})^4},
\end{equation}
also depends on its horizontal wavenumber $k_h$, horizontal phase velocity $c_i$ and the background wind $\bar{u}$. The other variables in (\ref{eq:molecular_dissipation}) are the kinematic viscosity $\mu_{mol}/\bar{\rho}$, the molecular viscosity for a given gas $\mu_{mol}$ and the Brunt-V\"ais\"al\"a frequency N, which is a function of temperature. 

It is seen that $\beta_{mol}$ is inversely proportional to the background density $\bar{\rho}$ and, therefore exponentially grows with height. Any GW harmonic surviving critical level filtering and propagating upward is eventually affected by molecular viscosity at a certain altitude, and ultimately ceases to exist. The parameters $\mu_{mol}$, $N$ and $\bar{\rho}$ depend on the atmospheric temperature $T$, which gives us a tool for explaining the observed day/night differences in GW activity.

\subsection{Nighttime and Daytime Gravity Wave Activity}\label{sec:grav-wave-activ-2}

\begin{figure*}[t!]
 \centering\hspace*{-1.cm}
 \includegraphics[width=0.7\textwidth]{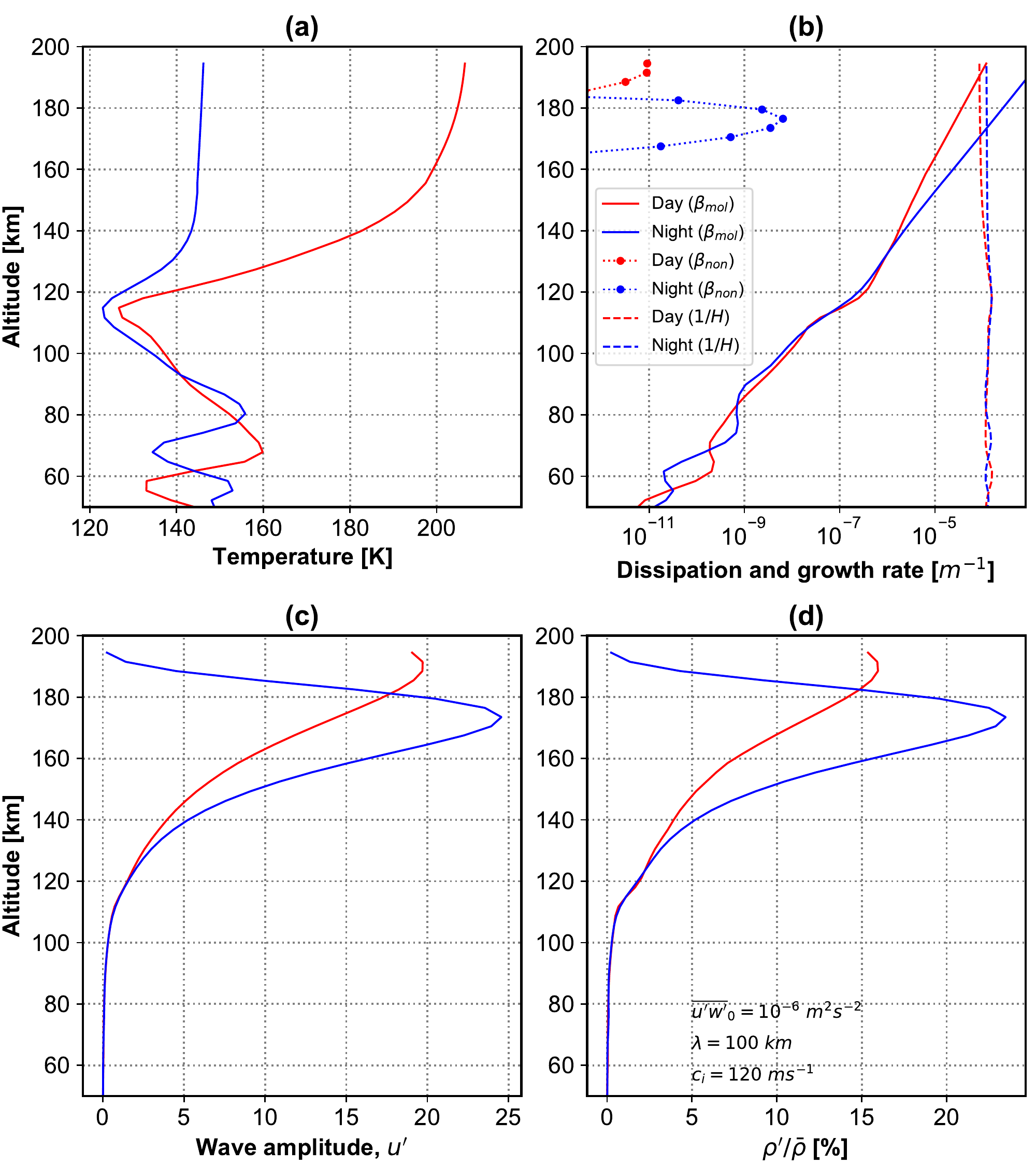}
 \caption{(a) Daytime (red) and nighttime (blue) Martian general circulation model output for neutral temperature in K, which is used as an input for the gravity wave (GW) model; (b) Vertical damping rate (dissipation) due to molecular viscosity ($\beta_{mol}$) and nonlinear diffusion $\beta_{non}$ compared with the wave growth rate $1/H$, (c) wave amplitude $u^\prime$, (d) GW-induced relative density perturbations $\rho^\prime/\bar{\rho}$. The profiles are for $45^\circ$N, $180^\circ$E, and $L_s = 90^\circ$ for solar minimum conditions. A single GW with a  representative horizontal wavelength of $\lambda=100$ km, horizontal phase speed $c_i = 120$ m s$^{-1}$, and initial flux of $\overline{u^\prime w^\prime}_0 = 10^{-6}$ m$^2$ s$^{-2}$ is used. }
 \label{fig:gw_model}
\end{figure*}

The MAVEN observations consistently show that the thermospheric GW activity is significantly larger during nighttime than daytime, as seen in the monthly mean analyses (Figures \ref{fig:GW_activity_solar_min}) as well as in the climatological mean of the GW activity during northern spring and summer seasons (Figures~\ref{fig:GW_activity_solar_min_ALT}-~\ref{fig:GW_activity_Lat_LT}). A day-night contrast has also been observed in previous studies of thermospheric GW activity based on MAVEN data, which focused on the altitude interval between 180-200 km and latitudes 62-70$^\circ$N, covering local times between 2-10 h during perihelion \citep{Yigit_etal15b}. \citet{Terada_etal17} have found that GW amplitudes clearly depend on the ambient atmospheric temperature. More recently, \citet{Siddle_etal19} used MAVEN observations up until August 2017 to study GW characteristics to show that GWs exhibit a trend with solar zenith angle (SZA), which is indicative of an inverse temperature dependence as well. \citet{Terada_etal17} and \citet{Siddle_etal19} suggested that the anticorrelation is caused by wave breaking  due to convective instability, which is in our case accounted for by $\beta_{non}$. This mechanism seems unlikely, because of the exponentially growing kinematic viscosity, which dominates in the upper thermosphere.

In order to test these hypotheses,
 we have used the whole atmosphere gravity wave model \citep{Yigit_etal08}, as described in Appendix \ref{sec:gravity-wave-column}, using the output from a Martian general circulation model \citep{Medvedev_etal13}. We adopted an empirical expression for the molecular viscosity of \citep{Hickey_etal15}
\begin{equation}
  \label{eq:molvis}
  \mu_{mol} = 4.5\times 10^5 \times (T/1000.0)^{0.71},
\end{equation}
which closely approximates molecular viscosity based on the kinetic theory. Typical vertical profiles of temperature for noon and midnight were taken from the model (up to 160 km) and extended smoothly higher up to 200 km to cover the altitudes, where the MAVEN observations took place. They are plotted in Figure~\ref{fig:gw_model}a and show that the day/night differences increase above the mesopause reaching $\sim$60 K at 200 km. The net temperature effect on $\beta_{mol}$ for a selected harmonic with the phase velocity $c_i=120$ m~s$^{-1}$, $k_h=2\pi/(100~\mathrm{km})$ and the incident flux of $\overline{u^\prime w^\prime}_0 = 10^{-6}$ m$^2$~s$^{-2}$, which are typical values for harmonics reaching the thermosphere \citep{Medvedev_etal11a}, is summarized in Figure~\ref{fig:gw_model}b. The  wave magnitudes (Figures~\ref{fig:gw_model}c,d) are determined by the balance between the growth rate associated with density stratification $1/H$ and damping rates $\beta$. The nighttime values exceed those during the day above a certain height ($\sim$130 km in this case). The growth rate is greater than $\beta_{mol}$ by several orders of magnitude up to $\sim$170 and 190 km at night and day, correspondingly, and the wave propagation below is mainly controlled by it. As a result, the amplitudes grow in vertical more steeper at midnight and reach maxima, where $\beta_{mol}$ approaches $1/H$. Nonlinear breaking/saturation was also included in the calculations, but $\beta_{non}$ is much smaller than $\beta_{mol}$ at these heights (Figure~\ref{fig:gw_model}b). Thus, the convective instability plays only a minor role in shaping the amplitudes of GWs in the thermosphere. The magnitudes of wave-induced horizontal wind fluctuations $u^\prime$ (Figure~\ref{fig:gw_model}b, $\sim$25 and 20 m~s$^{-1}$ for night and day, respectively) are far smaller than $\hat{c}_i$, also indicating that the onset of instability is nowhere near. In the end, the nighttime wave amplitude $\rho^\prime/\bar{\rho}$ exceeds the daytime value (Figure~\ref{fig:gw_model}d). The height of the amplitude maximum depends on the parameters of the harmonic, and increases with the phase velocity $c_i$. Our selection served for illustration purposes only.

\section{Summary and Conclusions}
\label{sec:summary-conclusions}
Using data from the NGIMS instrument on board NASA's MAVEN spacecraft, we have studied the climatology of thermospheric gravity wave (GW) activity during the recent extended solar minimum. For this, we analyzed the altitude, latitude, local time, and solar zenith angle dependencies of the GW-induced relative fluctuations of the carbon dioxide density in the Martian thermosphere from 150 to 230 km. We systematically quantified GW activity that encompassed month-to-month, day-to-day, and orbit-to-orbit variability.

The main findings of our research are as follows:
\begin{enumerate}
\item The thermosphere is pervaded by strong GW activity during the solar minimum. Their effects are continuously observed as small-scale density fluctuations with a significant amount of spatio-temporal variability, which is indicative of a broad spectrum of propagating thermospheric GWs.
\item Monthly mean GW activity varies strongly as a function of altitude (150--230 km) between 6-25\%, reaching a maximum at $\sim$170 km. 
\item Thermospheric GW activity is, generally, larger at middle- to high-latitudes. This is opposite to the lower atmosphere, where maxima are usually located at low-latitudes. This difference highlights the importance of wave filtering of short vertical-scale harmonics by the underlying winds, which is a manifestation of the vertical coupling between the lower and upper atmosphere on Mars. 
\item Thermospheric GW activity exhibits strong local time variations with nighttime magnitudes being larger than the daytime. Accordingly, GW activity increases with solar zenith angle, as was suggested by previous studies.
\item The day-night difference in the thermospheric GW activity is  the result of a competition between dissipation due to molecular diffusion and wave growth due to decreasing density, with the latter dominating the former up to a certain height in the upper thermosphere. Thus, the convective instability mechanism is likely less important in limiting the amplitude growth in the upper thermosphere and explaining the local time difference. 
\end{enumerate}

There is unambiguous observational evidence that GWs substantially affect the thermospheric variability. The thermosphere is the interface between the atmosphere and the space. It is the region where spacecraft perform aerobraking maneuvers and form their orbits. Atmospheric and ionospheric species escape to space in upper thermosphere,  while GWs can facilitate this process. Therefore, it is essential to better characterize GW processes in the Martian upper atmosphere and provide insight into their interaction with the large-scale  circulation, as has been done in this study. 

\section*{Acknowledgements}
E.Y. acknowledge support from NASA grant 19-MDAP19\_2-0105.
The NGIMS level 2, version 8 data supporting this article are publicly available at\\ \href{https://atmos.nmsu.edu/data_and_services/atmospheres_data/MAVEN/ngims.html}{https://atmos.nmsu.edu/data\_and\_services/atmospheres\_data/MAVEN/ngims.html}. The sunspot number data can be obtained from \href{http://www.sidc.be/silso/datafiles}{http://www.sidc.be/silso/datafiles}.
The Solar Dynamic Observatory images of the Sun can be obtained from \href{https://sdo.gsfc.nasa.gov/data/}{https://sdo.gsfc.nasa.gov/data/}.

\appendix
\section{Uncertainty in Gravity Wave Analysis}
\label{sec:uncert-grav-wave}

Our results have provided a climatology of gravity wave activity in terms of relative density fluctuations $\rho^\prime/\bar{\rho}$ by averaging over the retrieved $\rho^\prime/\bar{\rho}$ within a given bin. The number of measurements $N$ in a data bin is essential to the statistical significance of the retrieved GW activity values. In the different types of data bins presented in this study we typically have $N=50-15,000$. For example, in the monthly mean GW activity presented in Figure \ref{fig:GW_activity_solar_min}a, where the activity is studied as a function of height in 10 km bins for each month, we have $N=1437-15,246$ for each bin. The associated mean GW activity varies significantly, depending on the month and altitude. At 170 km (i.e., 170-180 km bin), it is 
\begin{equation}
  \overline{\rho^\prime/\bar{\rho}}(z=170~\mathrm{km}) \approx 24.6\%,
\end{equation}
and $N= 5536$. Accordingly, the \textit{standard mean error} is
\begin{equation}
  \sigma (\overline{\rho^\prime/\bar{\rho}}) \approx 0.0459\%.
\end{equation}
Therefore, in the representative bin where we have the maximum GW activity, the uncertainty of the GW activity is given by:
\begin{equation}
  \overline{\rho^\prime/\bar{\rho}} \approx 24.6 \pm 0.0459\%.
\end{equation}
The associated \textit{fractional uncertainty} is given by 
\begin{equation}
\frac{\sigma (\overline{\rho^\prime/\bar{\rho}})}{\overline{\rho^\prime/\bar{\rho}}}
 \approx 0.186\%.
\end{equation}
For the rest of the profiles presented in Figure~\ref{fig:GW_activity_solar_min}a, the range of fractional uncertainty is $\sim0.12\%-0.78\%$.

We have performed a similar uncertainty analysis for the rest of the observations of the mean GW activity, and the fractional uncertainty has been found not to exceed 1\%.

\section{Gravity Wave Model}
\label{sec:gravity-wave-column}
We used a whole atmosphere gravity wave (GW) parameterization \citep{Yigit_etal08} to study the upward propagation and dissipation of gravity waves in the Martian atmosphere. The gravity wave parameterization has been originally developed for an Earth general circulation model (GCM), the Coupled Middle Atmosphere-Thermosphere-2 GCM, to represent the sub-grid-scale gravity waves and has been used since in a number of Martian and Earth three-dimensional studies. The most recent applications for the terrestrial and Martian atmospheres can be found in the works by \citet{Yigit_etal21a} and \citet{Yigit_etal18}, respectively. The scheme calculates the vertical evolution of a broad-spectrum of monochromatic GWs prescribed  in terms of the wave momentum fluxes as a function of horizontal phase speeds \citep[][Figure 6]{YigitMedvedev10} at a source level in the lower atmosphere. In its application for Mars, the source level is between 5-8 km. The total vertical damping rate $\beta$ of a spectrum of waves is given by the superposition of the damping rates for each wave harmonic $i$  due to molecular viscosity $\beta_{mol}^i$ and nonlinear diffusion $\beta_{non}^i$:
\begin{equation}
  \label{eq:total_dissipation}
\beta = \sum_i \sum_d \beta_d^i = \sum_i \beta_{mol}^i + \beta_{non}^i,
\end{equation}
whose functional forms are given in detail in the work by \citet{Yigit_etal08}. While the GW model can be run as a stand-alone framework with appropriate input variables, such as temperature and density, it can be coupled with an atmospheric general circulation model to account for the effects of subgrid-scale GWs. It calculates the GW-induced mixing, energy and momentum deposition required in coarse-grid GCMs. The model makes steady-state and single-column approximations, where the individual harmonics propagate vertically upward. This assumption neglects wave reflection processes, which is well justified for the typically considered GW harmonics. The required input parameters for the scheme are the characteristic horizontal wavelength $\lambda_h = 2\pi/k_h$, initial wave fluxes $\overline{u^\prime w^\prime}_i$ and horizontal phase speeds $c_i$ for each wave harmonic, which are specified by an empirical GW source spectrum at the source level. No artificial or intermittancy factors are used in the model. 

\newpage

\begin{thebibliography}{}
\expandafter\ifx\csname natexlab\endcsname\relax\def\natexlab#1{#1}\fi
\providecommand{\url}[1]{\href{#1}{#1}}
\providecommand{\dodoi}[1]{doi:~\href{http://doi.org/#1}{\nolinkurl{#1}}}
\providecommand{\doeprint}[1]{\href{http://ascl.net/#1}{\nolinkurl{http://ascl.net/#1}}}
\providecommand{\doarXiv}[1]{\href{https://arxiv.org/abs/#1}{\nolinkurl{https://arxiv.org/abs/#1}}}

\bibitem[{Ando {et~al.}(2012)Ando, Imamura, \& Tsuda}]{Ando_etal12}
Ando, H., Imamura, T., \& Tsuda, T. 2012, J. Atmos. Sci., 69, 2906,
  \dodoi{10.1175/JAS-D-11-0339.1}

\bibitem[{Benna {et~al.}(2019)Benna, Bougher, Lee, Roeten, Yi\u{g}it, Mahaffy,
  \& Jakosky}]{Benna_etal19}
Benna, M., Bougher, S.~W., Lee, Y., {et~al.} 2019, Science, 366, 1363,
  \dodoi{10.1126/science.aax1553}

\bibitem[{Bougher {et~al.}(2015{\natexlab{a}})Bougher, Cravens, Grebowsky, \&
  Luhmann}]{Bougher_etal15b}
Bougher, S.~W., Cravens, T.~E., Grebowsky, J., \& Luhmann, J.
  2015{\natexlab{a}}, Space Sci. Rev., 195, 423,
  \dodoi{10.1007/s11214-014-0053-7}

\bibitem[{Bougher {et~al.}(1999)Bougher, Engel, Roble, \&
  Foster}]{Bougher_etal99}
Bougher, S.~W., Engel, S., Roble, R.~G., \& Foster, B. 1999, J. Geophys. Res.,
  104, 16,591, \dodoi{10.1029/1998JE001019}

\bibitem[{Bougher {et~al.}(2015{\natexlab{b}})Bougher, Pawlowski, Bell, Nelli,
  McDunn, Murphy, Chizek, \& Ridley}]{Bougher_etal15}
Bougher, S.~W., Pawlowski, D., Bell, J.~M., {et~al.} 2015{\natexlab{b}}, J.
  Geophys. Res. Planets, 120, 311, \dodoi{10.1002/2014JE004715}

\bibitem[{Bougher {et~al.}(2017)Bougher, Roeten, Olsen, Mahaffy, Benna, Elrod,
  Jain, Schneider, Deighan, Thiemann, Eparvier, Stiepen, \&
  Jakosky}]{Bougher_etal17}
Bougher, S.~W., Roeten, K.~J., Olsen, K., {et~al.} 2017, J. Geophys. Res. Space
  Physics, 122, 1296, \dodoi{https://doi.org/10.1002/2016JA023454}

\bibitem[{Chanin(2007)}]{Chanin07}
Chanin, M.~L. 2007, in Solar {Variability} and {Planetary} {Climates}, ed.
  Y.~Calisesi, R.~M. Bonnet, L.~Gray, J.~Langen, \& M.~Lockwood, Space
  {Sciences} {Series} of {ISSI} (Springer), 261--272,
  \dodoi{10.1007/978-0-387-48341-2\_21}

\bibitem[{Creasey {et~al.}(2006{\natexlab{a}})Creasey, Forbes, \&
  Hinson}]{Creasey_etal06a}
Creasey, J.~E., Forbes, J.~M., \& Hinson, D.~P. 2006{\natexlab{a}}, Geophys.
  Res. Lett., 33, \dodoi{10.1029/2005GL024037}

\bibitem[{Creasey {et~al.}(2006{\natexlab{b}})Creasey, Forbes, \&
  Keating}]{Creasey_etal06b}
Creasey, J.~E., Forbes, J.~M., \& Keating, G.~M. 2006{\natexlab{b}}, Geophys.
  Res. Lett., 33, \dodoi{10.1029/2006GL027583}

\bibitem[{Dikpati {et~al.}(2004)Dikpati, Toma, Gilman, Arge, \&
  White}]{Dikpati_etal04}
Dikpati, M., Toma, G.~d., Gilman, P.~A., Arge, C.~N., \& White, O.~R. 2004,
  Astrophys. J., 601, 1136, \dodoi{10.1086/380508}

\bibitem[{England {et~al.}(2017)England, Liu., Yi\u{g}it, Mahaffy, Elrod,
  Benna, Nakagawa, Terada, \& Jakosky}]{England_etal17}
England, S.~L., Liu., G., Yi\u{g}it, E., {et~al.} 2017, J. Geophys. Res. Space
  Physics, 2310–2335, \dodoi{10.1002/2016JA023475}

\bibitem[{Forbes {et~al.}(2016)Forbes, Bruinsma, Doornbos, \&
  Zhang}]{Forbes_etal16}
Forbes, J.~M., Bruinsma, S.~L., Doornbos, E., \& Zhang, X. 2016, J. Geophys.
  Res. Space Physics, 121, 6914, \dodoi{10.1002/2016JA022923}

\bibitem[{Fritts {et~al.}(2015)Fritts, Laughman, Lund, \&
  Snively}]{Fritts_etal15a}
Fritts, D.~C., Laughman, B., Lund, T.~S., \& Snively, J.~B. 2015, J. Geophys.
  Res. Atmos., 120, 8783, \dodoi{10.1002/2015JD023363}

\bibitem[{Fritts {et~al.}(2006)Fritts, Wang, \& Tolson}]{Fritts_etal06}
Fritts, D.~C., Wang, L., \& Tolson, R.~H. 2006, J. Geophys. Res., 111,
  \dodoi{10.1029/2006JA011897}

\bibitem[{Garcia {et~al.}(2009)Garcia, Drossart, Piccioni, López‐Valverde,
  \& Occhipinti}]{Garcia_etal09}
Garcia, R.~F., Drossart, P., Piccioni, G., López‐Valverde, M., \&
  Occhipinti, G. 2009, J. Geophys. Res. Planets, 114,
  \dodoi{10.1029/2008JE003073}

\bibitem[{Gavrilov \& Kshevetskii(2013)}]{GavrilovKshevetskii13}
Gavrilov, N.~M., \& Kshevetskii, S.~P. 2013, Adv. Space Res., 51, 1168,
  \dodoi{10.1016/j.asr.2012.10.023}

\bibitem[{Gavrilov \& Kshevetskii(2015)}]{GavrilovKshevetskii15}
---. 2015, Advances in Space Research, 56, 1833,
  \dodoi{http://dx.DOI.org/10.1016/j.asr.2015.01.033}

\bibitem[{Gavrilov {et~al.}(2020)Gavrilov, Kshevetskii, \&
  Koval}]{Gavrilov_etal20}
Gavrilov, N.~M., Kshevetskii, S.~P., \& Koval, A.~V. 2020, J. Atmos. Sol.-Terr.
  Phys., 105381, \dodoi{10.1016/j.jastp.2020.105381}

\bibitem[{González‐Galindo {et~al.}(2015)González‐Galindo,
  López‐Valverde, Forget, García‐Comas, Millour, \&
  Montabone}]{Gonzalez-Galindo_etal15}
González‐Galindo, F., López‐Valverde, M.~A., Forget, F., {et~al.} 2015,
  J. Geophys. Res. Planets, 120, 2020,
  \dodoi{https://doi.org/10.1002/2015JE004925}

\bibitem[{Heale {et~al.}(2014)Heale, Snively, Hickey, \& Ali}]{Heale_etal14}
Heale, C.~J., Snively, J.~B., Hickey, M.~P., \& Ali, C.~J. 2014, J. Geophys.
  Res. Space Physics, 119, 3857, \dodoi{10.1002/2013JA019387}

\bibitem[{Heavens {et~al.}(2020)Heavens, Kass, Kleinb\"ohl, \&
  Schofield}]{Heavens_etal20a}
Heavens, N., Kass, D.~M., Kleinb\"ohl, A., \& Schofield, J.~T. 2020, Icarus,
  341, 113630, \dodoi{10.1016/j.icarus.2020.113630}

\bibitem[{Hickey \& Cole(1988)}]{HickeyCole88}
Hickey, M.~P., \& Cole, K.~D. 1988, J. Atmos. Terr. Phys., 50, 689

\bibitem[{Hickey {et~al.}(2011)Hickey, Walterscheid, \&
  Schubert}]{Hickey_etal11}
Hickey, M.~P., Walterscheid, R.~L., \& Schubert, G. 2011, J. Geophys. Res.,
  116, \dodoi{10.1029/2011JA016792}

\bibitem[{Hickey {et~al.}(2015)Hickey, Walterscheid, \&
  Schubert}]{Hickey_etal15}
---. 2015, J. Geophys. Res. Space Physics, 120, 3074,
  \dodoi{10.1002/2014JA020583}

\bibitem[{Jain {et~al.}(2020)Jain, Bougher, Deighan, Schneider,
  González Galindo, Stewart, Sharrar, Kass, Murphy, \&
  Pawlowski}]{Jain_etal20}
Jain, S.~K., Bougher, S.~W., Deighan, J., {et~al.} 2020, Geophys. Res. Lett.,
  47, \dodoi{10.1029/2019GL085302}

\bibitem[{Jesch {et~al.}(2019)Jesch, Medvedev, Castellini, Yi\u{g}it, \&
  Hartogh}]{Jesch_etal19}
Jesch, D., Medvedev, A.~S., Castellini, F., Yi\u{g}it, E., \& Hartogh, P. 2019,
  Atmosphere, 10, 620, \dodoi{10.3390/atmos10100620}

\bibitem[{Jiang {et~al.}(2021)Jiang, Yokoyama, Wei, Yang, \&
  Zhao}]{Jiang_etal21}
Jiang, C., Yokoyama, T., Wei, L., Yang, G., \& Zhao, Z. 2021, Astrophys. J.,
  909, 47, \dodoi{10.3847/1538-4357/abdc1d}

\bibitem[{Kiess {et~al.}(2014)Kiess, Reza, \& Wolfgang}]{Kiess_etal14}
Kiess, C., Reza, R., \& Wolfgang, S. 2014, Astron. Astrophys., 10,
  \dodoi{10.1051/0004-6361/201321119}

\bibitem[{Kilcik {et~al.}(2011)Kilcik, Yurchyshyn, Abramenko, Goode, Ozguc,
  Rozelot, \& Cao}]{Kilcik_etal11}
Kilcik, A., Yurchyshyn, V.~B., Abramenko, V., {et~al.} 2011, Astrophys. J.,
  731, \dodoi{10.1088/0004-637X/731/1/30}

\bibitem[{Kshevetskii \& Gavrilov(2005)}]{KshevetskiiGavrilov05}
Kshevetskii, S.~P., \& Gavrilov, N.~M. 2005, J. Atmos. Sol.-Terr. Phys., 67,
  1014 , \dodoi{http://dx.DOI.org/10.1016/j.jastp.2005.02.013}

\bibitem[{Kuroda {et~al.}(2020)Kuroda, Medvedev, \& Yiğit}]{Kuroda_etal20}
Kuroda, T., Medvedev, A.~S., \& Yiğit, E. 2020, J. Geophys. Res. Planets, 125,
  \dodoi{10.1029/2020JE006556}

\bibitem[{Kuroda {et~al.}(2019)Kuroda, Yiğit, \& Medvedev}]{Kuroda_etal19}
Kuroda, T., Yiğit, E., \& Medvedev, A.~S. 2019, J. Geophys. Res. Planets, 124,
  1618, \dodoi{10.1029/2018JE005847}

\bibitem[{Kutiev {et~al.}(2013)Kutiev, Tsagouri, Perrone, Pancheva, Mukhtarov,
  Mikhailov, Lastovicka, Jakowski, Buresova, Blanch, Andonov, Altadill,
  Magdaleno, Parisi, \& Torta}]{Kutiev_etal13}
Kutiev, I., Tsagouri, I., Perrone, L., {et~al.} 2013, Journal of Space Weather
  and Space Climate, 3, A06, \dodoi{10.1051/swsc/2013028}

\bibitem[{Leelavathi {et~al.}(2020)Leelavathi, Venkateswara~Rao, \&
  Rao}]{Leelavathi_etal20}
Leelavathi, V., Venkateswara~Rao, N., \& Rao, S. V.~B. 2020, J. Geophys. Res.
  Planets, 125, e2020JE006649, \dodoi{10.1029/2020JE006649}

\bibitem[{Li {et~al.}(2019)Li, Xiang, Xie, \& Xu}]{Li_etal19}
Li, F.~Y., Xiang, N.~B., Xie, J.~L., \& Xu, J.~C. 2019, Astrophys. J., 873,
  122, \dodoi{10.3847/1538-4357/ab06bf}

\bibitem[{Li {et~al.}(2021)Li, Liu, \& Jin}]{Li_etal21}
Li, Y., Liu, J., \& Jin, S. 2021, J. Geophys. Res. Space Physics, 126,
  e2020JA028378, \dodoi{10.1029/2020JA028378}

\bibitem[{Lilienthal {et~al.}(2020)Lilienthal, Yi\u{g}it, Samtleben, \&
  Jacobi}]{Lilienthal_etal20}
Lilienthal, F., Yi\u{g}it, E., Samtleben, N., \& Jacobi, C. 2020, Front.
  Astron. Space Sci., 2020, 7:588956, \dodoi{10.3389/fspas.2020.588956}

\bibitem[{Lindzen(1981)}]{Lindzen81}
Lindzen, R.~S. 1981, J. Geophys. Res., 86, 9707

\bibitem[{Lund \& Fritts(2012)}]{LundFritts12}
Lund, T.~S., \& Fritts, D.~C. 2012, J. Geophys. Res., 117,
  \dodoi{10.1029/2012JD017536, 2012}

\bibitem[{Mahaffy {et~al.}(2015)Mahaffy, Benna, Elrod, Yelle, Bougher, Stone,
  \& Jakosky}]{Mahaffy_etal15}
Mahaffy, P.~R., Benna, M., Elrod, M., {et~al.} 2015, Geophys. Res. Lett., 42,
  8951, \dodoi{10.1002/2015GL065329}

\bibitem[{Mayyasi {et~al.}(2019)Mayyasi, Narvaez, Benna, Elrod, \&
  Mahaffy}]{Mayyasi_etal19}
Mayyasi, M., Narvaez, C., Benna, M., Elrod, M., \& Mahaffy, P. 2019, J.
  Geophys. Res. Space Physics, 124,
  \dodoi{https://doi.org/10.1029/2019JA026481}

\bibitem[{Medvedev {et~al.}(2015)Medvedev, Gonz\'{a}lez-Galindo, Yi\u{g}it,
  Feofilov, Forget, \& Hartogh}]{Medvedev_etal15}
Medvedev, A.~S., Gonz\'{a}lez-Galindo, F., Yi\u{g}it, E., {et~al.} 2015, J.
  Geophys. Res. Planets, 120, 913, \dodoi{10.1002/2015JE004802}

\bibitem[{Medvedev \& Klaassen(1995)}]{MedvedevKlaassen95}
Medvedev, A.~S., \& Klaassen, G.~P. 1995, J. Geophys. Res., 100, 25841

\bibitem[{Medvedev \& Klaassen(2000)}]{MedvedevKlaassen00}
---. 2000, J. Atmos. Sol.-Terr. Phys., 62, 1015

\bibitem[{Medvedev \& Yi\u{g}it(2012)}]{MedvedevYigit12}
Medvedev, A.~S., \& Yi\u{g}it, E. 2012, Geophys. Res. Lett., 39,
  \dodoi{10.1029/2012GL050852}

\bibitem[{Medvedev \& Yi\u{g}it(2019)}]{MedvedevYigit19}
---. 2019, Atmosphere, 10, \dodoi{10.3390/atmos10090531}

\bibitem[{Medvedev {et~al.}(2011)Medvedev, Yi\u{g}it, \&
  Hartogh}]{Medvedev_etal11a}
Medvedev, A.~S., Yi\u{g}it, E., \& Hartogh, P. 2011, Icarus, 211, 909,
  \dodoi{10.1016/j.icarus.2010.10.013}

\bibitem[{Medvedev {et~al.}(2013)Medvedev, Yi\u{g}it, Kuroda, \&
  Hartogh}]{Medvedev_etal13}
Medvedev, A.~S., Yi\u{g}it, E., Kuroda, T., \& Hartogh, P. 2013, J. Geophys.
  Res. Planets, 118, 1, \dodoi{10.1002/jgre.20163, 2013}

\bibitem[{Miyoshi \& Yi\u{g}it(2019)}]{MiyoshiYigit19}
Miyoshi, Y., \& Yi\u{g}it, E. 2019, Ann. Geophys., 37, 955,
  \dodoi{10.5194/angeo-37-955-2019}

\bibitem[{Müller‐Wodarg {et~al.}(2019)Müller‐Wodarg, Koskinen, Moore,
  Serigano, Yelle, Hörst, Waite, \& Mendillo}]{Muller-Wodarg_etal19}
Müller‐Wodarg, I. C.~F., Koskinen, T.~T., Moore, L., {et~al.} 2019, Geophys.
  Res. Lett., 46, 2372, \dodoi{10.1029/2018GL081124}

\bibitem[{Parish {et~al.}(2009)Parish, Schubert, Hickey, \&
  Walterscheid}]{Parish_etal09}
Parish, H.~F., Schubert, G., Hickey, M., \& Walterscheid, R.~L. 2009, Icarus,
  203, 28

\bibitem[{Park {et~al.}(2014)Park, L\"uhr, Lee, Kim, Jee, \& Kim}]{Park_etal14}
Park, J., L\"uhr, H., Lee, C., {et~al.} 2014, J. Geophys. Res. Space Physics,
  119, \dodoi{10.1002/2013JA019705}

\bibitem[{Pesnell(2020)}]{Pesnell20}
Pesnell, W.~D. 2020, J. Space Weather Space Clim., 10, 60,
  \dodoi{10.1051/swsc/2020060}

\bibitem[{Siddle {et~al.}(2019)Siddle, Mueller-Wodarg, Stone, \&
  Yelle}]{Siddle_etal19}
Siddle, A., Mueller-Wodarg, I., Stone, S., \& Yelle, R. 2019, Icarus, 333, 12,
  \dodoi{10.1016/j.icarus.2019.05.021}

\bibitem[{Straus {et~al.}(2008)Straus, Fleck, Jefferies, Cauzzi, McIntosh,
  Reardon, Severino, \& Steffen}]{Straus_etal08}
Straus, T., Fleck, B., Jefferies, S.~M., {et~al.} 2008, Astrophys. J., 681,
  L125, \dodoi{10.1086/590495}

\bibitem[{Svalgaard {et~al.}(2005)Svalgaard, Cliver, \&
  Kamide}]{Svalgaard_etal05}
Svalgaard, L., Cliver, E.~W., \& Kamide, Y. 2005, Geophys. Res. Lett., 32,
  \dodoi{https://doi.org/10.1029/2004GL021664}

\bibitem[{Terada {et~al.}(2017)Terada, Leblanc, Nakagawa, Medvedev, Yi\u{g}it,
  Kuroda, Hara, England, Fujiwara, Terada, Seki, Mahaffy, Elrod, Benna,
  Grebowsky, \& Jakosky}]{Terada_etal17}
Terada, N., Leblanc, F., Nakagawa, H., {et~al.} 2017, J. Geophys. Res. Space
  Physics, \dodoi{10.1002/2016JA023476}

\bibitem[{Thiemann {et~al.}(2018)Thiemann, Andersson, Lillis, Withers, Xu,
  Elrod, Jain, Pilinski, Pawlowski, Chamberlin, Eparvier, Benna, Fowler, Curry,
  Peterson, \& Deighan}]{Thiemann_etal18}
Thiemann, E. M.~B., Andersson, L., Lillis, R., {et~al.} 2018, Geophys. Res.
  Lett., 45, 8005, \dodoi{10.1029/2018GL077730}

\bibitem[{Tolson {et~al.}(2007)Tolson, Keating, Zurek, Bougher, Justus, \&
  Fritts}]{Tolson_etal07}
Tolson, R., Keating, G., Zurek, R.~W., {et~al.} 2007, J. Spacecraft Rockets,
  44, 1172

\bibitem[{Vals {et~al.}(2019)Vals, Spiga, Forget, Millour, Montabone, \&
  Lott}]{Vals_etal19}
Vals, M., Spiga, A., Forget, F., {et~al.} 2019, Planet. Space Sci., 178,
  104708, \dodoi{10.1016/j.pss.2019.104708}

\bibitem[{Vickers {et~al.}(2014)Vickers, Kosch, Sutton, Bjoland, Ogawa, \&
  La~Hoz}]{Vickers_etal14}
Vickers, H., Kosch, M.~J., Sutton, E., {et~al.} 2014, J. Geophys. Res. Space
  Physics, 119, 6833, \dodoi{10.1002/2014JA019885}

\bibitem[{Vigren \& Cui(2019)}]{VigrenCui19}
Vigren, E., \& Cui, J. 2019, Astrophys. J., 887, 177,
  \dodoi{10.3847/1538-4357/ab53db}

\bibitem[{Walterscheid \& Hickey(2011)}]{WalterscheidHickey11}
Walterscheid, R.~L., \& Hickey, M.~P. 2011, J. Geophys. Res., 116, D12101,
  \dodoi{10.1029/2010JD014987}

\bibitem[{Walterscheid {et~al.}(2013)Walterscheid, Hickey, \&
  Schubert}]{Walterscheid_etal13}
Walterscheid, R.~L., Hickey, M.~P., \& Schubert, G. 2013, J. Geophys. Res.
  Planets, 118, 2169, \dodoi{10.1002/jgre.20164}

\bibitem[{Watkins \& Cho(2010)}]{WatkinsCho10}
Watkins, C., \& Cho, J. Y.-K. 2010, Astrophys. J., 714, 904,
  \dodoi{10.1088/0004-637X/714/1/904}

\bibitem[{Watkins \& Cho(2013)}]{WatkinsCho13}
---. 2013, Geophysical Research Letters, 40, 472, \dodoi{10.1029/2012GL054368}

\bibitem[{Yi\u{g}it(2018)}]{Yigit18}
Yi\u{g}it, E. 2018, Atmospheric and Space Sciences: Ionospheres and Plasma
  Environments (Volume 2), SpringerBriefs in Earth Sci. (Springer,
  Netherlands), \dodoi{10.1007/978-3-319-62006-0}

\bibitem[{Yi\u{g}it {et~al.}(2021{\natexlab{a}})Yi\u{g}it, Alexander, \&
  Ern}]{Yigit_etal21a}
Yi\u{g}it, E., Alexander, A.~S., \& Ern, M. 2021{\natexlab{a}}, Front. Astron.
  Space Sci., 7, \dodoi{10.3389/fspas.2020.614018}

\bibitem[{Yi\u{g}it {et~al.}(2008)Yi\u{g}it, Aylward, \&
  Medvedev}]{Yigit_etal08}
Yi\u{g}it, E., Aylward, A.~D., \& Medvedev, A.~S. 2008, J. Geophys. Res., 113,
  \dodoi{10.1029/2008JD010135}

\bibitem[{Yi\u{g}it {et~al.}(2015{\natexlab{a}})Yi\u{g}it, England, Liu,
  Medvedev, Mahaffy, Kuroda, \& Jakowsky}]{Yigit_etal15b}
Yi\u{g}it, E., England, S.~L., Liu, G., {et~al.} 2015{\natexlab{a}}, Geophys.
  Res. Lett., 42, \dodoi{10.1002/2015GL065307}

\bibitem[{Yi\u{g}it {et~al.}(2018{\natexlab{a}})Yi\u{g}it, Kilcik, Elias,
  Dönmez, Ozguc, Yurchshyn, \& Rozelot}]{Yigit_etal18a}
Yi\u{g}it, E., Kilcik, A., Elias, A.~G., {et~al.} 2018{\natexlab{a}}, J. Atmos.
  Sol.-Terr. Phys., 171, 157, \dodoi{10.1016/j.jastp.2017.11.018}

\bibitem[{Yi\u{g}it {et~al.}(2016)Yi\u{g}it, Kn\'i\v{z}ov\'{a}, Georgieva, \&
  Ward}]{Yigit_etal16b}
Yi\u{g}it, E., Kn\'i\v{z}ov\'{a}, P.~K., Georgieva, K., \& Ward, W. 2016, J.
  Atmos. Sol.-Terr. Phys., 141, 1,
  \dodoi{http://dx.DOI.org/10.1016/j.jastp.2016.02.011}

\bibitem[{Yi\u{g}it \& Medvedev(2010)}]{YigitMedvedev10}
Yi\u{g}it, E., \& Medvedev, A.~S. 2010, J. Geophys. Res., 115,
  \dodoi{10.1029/2009JA015106}

\bibitem[{Yi\u{g}it \& Medvedev(2017)}]{YigitMedvedev17}
---. 2017, J. Geophys. Res. Space Physics, 122, 4846–4864,
  \dodoi{10.1002/2017JA024089}

\bibitem[{Yi\u{g}it \& Medvedev(2019)}]{YigitMedvedev19}
---. 2019, Physics Today, 6, 40, \dodoi{10.1063/PT.3.4226}

\bibitem[{Yi\u{g}it {et~al.}(2021{\natexlab{b}})Yi\u{g}it, Medvedev, Benna, \&
  Jakosky}]{Yigit_etal21b}
Yi\u{g}it, E., Medvedev, A.~S., Benna, M., \& Jakosky, B. 2021{\natexlab{b}},
  Geophys. Res. Lett., 48, e2020GL092095, \dodoi{10.1029/2020GL092095}

\bibitem[{Yi\u{g}it {et~al.}(2015{\natexlab{b}})Yi\u{g}it, Medvedev, \&
  Hartogh}]{Yigit_etal15a}
Yi\u{g}it, E., Medvedev, A.~S., \& Hartogh, P. 2015{\natexlab{b}}, Geophys.
  Res. Lett., 42, 4294, \dodoi{10.1002/2015GL064275}

\bibitem[{Yi\u{g}it {et~al.}(2018{\natexlab{b}})Yi\u{g}it, Medvedev, \&
  Hartogh}]{Yigit_etal18}
---. 2018{\natexlab{b}}, Ann. Geophys., 36, 1631,
  \dodoi{10.5194/angeo-36-1631-2018}

\bibitem[{Young {et~al.}(1997)Young, Yelle, Young, Seiff, \&
  Kirk}]{Young_etal97}
Young, L.~A., Yelle, R.~V., Young, R., Seiff, A., \& Kirk, D.~B. 1997, Science,
  276, 108, \dodoi{10.1126/science.276.5309.108}

\bibitem[{Zurek {et~al.}(2017)Zurek, Tolson, Bougher, Lugo, Baird, Bell, \&
  Jakosky}]{Zurek_etal17}
Zurek, R.~W., Tolson, R.~A., Bougher, S.~W., {et~al.} 2017, J. Geophys. Res.
  Space Physics, 122, \dodoi{10.1002/2016JA023641}

\end{thebibliography}

\end{document}